\documentclass[letterpaper,preprintnumbers,prd,nofootinbib,nobibnotes,showpacs]{revtex4}
\usepackage{amsfonts}
\usepackage{mathrsfs}
\usepackage{epsfig}
\usepackage{graphicx}%
\usepackage{dcolumn}
\usepackage{amsmath}
\usepackage{color}
\usepackage{subfigure}

\makeatletter
\def\btt#1{\texttt{\@backslashchar#1}}%
\DeclareRobustCommand\bblash{\btt{\@backslashchar}}%
\makeatother
\begin{document}

\title{Nonsingular black holes and nonsingular universes in the regularized Lovelock gravity}
\author{Changjun Gao}\email{gaocj@bao.ac.cn} \author{Shuang Yu}\email{yushuang@nao.cas.cn} \author{Jianhui Qiu}\email{jhqiu@nao.cas.cn}\affiliation{ Key Laboratory of Computational Astrophysics, National Astronomical Observatories, Chinese
Academy of Sciences, Beijing 100012, China}
\affiliation{University of Chinese Academy of Sciences, Beijing 100049, China}

\date{\today}

\begin{abstract}
It is found that, when the coupling constants $\alpha_p$ in the theory of regularized Lovelock gravity are properly chosen and the number of Lovelock tensors $p\rightarrow \infty$, there exist a fairly large number of nonsingular (singularity free) black holes and nonsingular universes. Some nonsingular black holes have numerous horizons and numerous energy levels (a bit like atom) inside the outer event horizon. On the other hand, some nonsingular universes start and end in two de Sitter phases. The ratio of energy densities for the two phases are $120$ orders. It is thus helpful to understand the cosmological constant problem.
\end{abstract}

\pacs{04.50.Kd,04.70.Bw,04.30.-w,04.80.Cc}

\maketitle

\section{Introduction}
The Lovelock theorem \cite{ll:1971} states that, in higher dimensions, General Relativity is not the unique healthy theory that has the second
order equation of motion and consists of the metric tensor together with its derivatives (up to second order). Actually, the most general gravitational theory leading to second order field equations and consisting of only
the metric tensor and its second derivatives in higher dimensions is the Lovelock gravity. When the dimension of spacetime is four, the Lovelock theory reduces to General Relativity \cite{ll:1972} due to the fact that the higher order Lovelock
invariants become total derivative. In view of this point, many attempts has been made in the literature to let the
higher order Lovelock tensors contribute in four dimensions. These include
the introduction of extra degrees of freedom, non-minimally coupled to the Lovelock invariants
\cite{nonminimal}, or the consideration of non-linear function constructed with the Lovelock
invariants \cite{non-l}.

Recently, a novel way is introduced in order to make the Lovelock tensors non-vanishing in four dimensional spacetime \cite{glavan:2020,CasalinoCRV}. (See also \cite{tom:2011,cog:2013} for earlier works). The key idea of the novel method amounts to regularize the coupling constants
\begin{eqnarray}
\alpha_p\rightarrow \tilde{\alpha}_p=\alpha_p\frac{\left(n-2p-1\right)!}{\left(n-1\right)!}\;.
\end{eqnarray}
Here $p$ and  $\alpha_p$ are the order and coupling constants in Lovelock invariants. $n$ is the dimension of spacetime. With this regularization, the Lovelock gravity becomes the \emph{regularized Lovelock gravity} \cite{glavan:2020,CasalinoCRV}. Then it is found the corresponding regularized Lovelock tensors are non-vanishing even if the spacetime is four dimensional. With the invention of regularized Lovelock gravity \cite{glavan:2020,CasalinoCRV}, many interesting discoveries are made. These discoveries are included in the new exact solutions \cite{KumarGhoshMaharaj,SinghGM,DonevaYazadjiev,JusufiBG,GeSin,Liu14267,Yang14468,MaLu}, the black hole quasinormal modes \cite{KonoplyaZinhailoZhidenko,Churilova,Mishra,LiWY,ZhangZLG,AragonBGV,MalafarinaTD,Cuyubamba09025,LiuNZ,Devi14935}, the black hole shadows \cite{GuoLi,WeiLiu07769,ZhangWeiLiu,Heydari-Fard,RayimbaevATA,ZengZZ}, the gravitational lensing \cite{LiuZW,KumaraRHAA,IslamKG,JinGL,Kumar12970}, the black hole thermodynamics \cite{HegdeKA,SinghS,ZhangLG,Mansoori,WeiL14275,Konoplya02248,PanahJafarzade,YangWCYW,Ying}, the regularized Einstein-Gauss-Bonnet gravity \cite{LuPangMao,Kobayashi,Fernandes08362,HennigarKMP,BonifacioHJ} and the holographic superconductors \cite{qiao}. Of course, there remain some objections on this theory \cite{Ai2020,GursesST,Mahapatra2020,Shu09339,TianZhu,ArrecheaDJ}. For example, by considering a quantum tunneling of vacua in the regularized four dimensional Einstein-Gauss-Bonnet gravity, Shu \cite{Shu09339} finds a disastrous divergence of vacuum decay rate. Thus an inconsistence of the theory is put forward. Putting aside the debates, we report in this paper that a large number of nonsingular black holes and nonsingular universes can be constructed in the regularized Lovelock gravity without introducing any physical sources.

Generally, there are two ways to obtain nonsingular black holes and nonsingular universes. The first way is to modify the right hand of Einstein equation and the second way is to modify the left hand. The first way aims at resorting to exotic physical sources and the second way amounts to modify gravity itself.
The nonsingular black hole models with various physical sources can be found in \cite{bh-5,bh-6,bh-7,bh-8,bh-9,bh-10,bh-11,bh-12,bh-13,bh-14,bh-15,bh-16,bh-17,bh-18,hayward:2006}. These nonsingular black holes have an event horizon and no singularities. Although they are derived in the framework of General Relativity, they avoid the well-known singularity theorems because their physical sources violate the energy conditions. The corresponding physical sources are mostly built with some non-linear electrodynamics. The other nonsingular black hole models are proposed in exact conformal field theory \cite{bh-19}, the noncommutative geometry \cite{nicolini:2006,nicolini:2007}, the string theory \cite{tsey:1995} and very recently, the Euler-Heisenberg theory of electrodynamics coupled to modified gravity \cite{gue:2020}. These black holes have often been referred to as ¡°Bardeen black holes¡± \cite{bh-9} because it is Bardeen that was the first one proposing the nonsingular black hole. The nonsingular universe models can be found in \cite{uni-1,uni-2,uni-3,lag-4,bran-5,report-6}. They are based on various approaches such as modified gravity models \cite{uni-1,uni-2}, Lagrangian multiplier
gravity actions (see e.g., \cite{lag-4}), brane world scenarios \cite{bran-5} an so on. Here we do not want to present an exhaustive list
of references, but we prefer the readers to read the nice
review paper by Novello and Bergliaffa \cite{report-6} and the references
therein.

The paper is organized as follows. In Sec. II, we review the theory of Lovelock gravity in $n$ dimensions. In Sec. III, we look for the four dimensional nonsingular black holes in the regularized Lovelock gravity. In order to make a connection with observations, we calculate the black hole quasinormal modes In Sec. IV. In order to understand the internal structure of the black holes, we investigate the geodesic motions of massless and massive particles in Sec. V. In Sec. VI, we look for the four dimensional nonsingular universes. Finally, Sec. VII gives the conclusion and discussion. Throughout this paper, we adopt the system of units in which $G=c=\hbar=1$ and the metric signature
$(-,\ +,\ +,\ +)$.

\section{The Lovelock gravity in n dimensions}\label{sec:2}
The action of Lovelock gravity \cite{ll:1971} in $n$ dimensions takes the form
\begin{eqnarray}
{S}=\int d^n x\sqrt{-g}\left(\sum_{p=0}^{N}\alpha_p L_{p}+L_m\right)\;,\label{ll}
\end{eqnarray}
where $n$ is the dimension of spacetime, $\alpha_p$ are dimensional constants and
summation is carried over all $p\in N$ with $N\leq (n-1)/2$. $L_m$ is the Lagrangian of matters.

$L_{p}$ is defined by
\begin{eqnarray}
L_{p}=2^{-p}\delta_{\sigma_1\sigma_2\cdot\cdot\cdot \sigma_{2p}}^{\lambda_1\lambda_2\cdot\cdot\cdot \lambda_{2p}}
R_{\lambda_1\lambda_2}^{\sigma_1\sigma_2}R_{\lambda_3\lambda_4}^{\sigma_3\sigma_4}\cdot\cdot\cdot R_{\lambda_{2p-1}\lambda_{2p}}^{\sigma_{2p-1}\sigma_{2p}}
\;,
\end{eqnarray}
where $\delta_{\sigma_1\sigma_2\cdot\cdot\cdot \sigma_{2p}}^{\lambda_1\lambda_2\cdot\cdot\cdot \lambda_{2p}}$
is the generalized Kronecker delta function of the order $2p$. It equals to $\pm 1$ if the upper indices form
an even or odd permutation of the lower ones, respectively,
and zero in all other cases. Here $R_{\lambda_i\lambda_j}^{\sigma_k\sigma_l}$ is the Riemann tensor. Since $L_p$ has the dimension of $l^{-2p}$, $\alpha_p$ has the dimension of $l^{2p-2}$ ($l$ is some length.).

For some specific examples, we have
\begin{eqnarray}
L_{0}=1\;,\ \ \  L_{1}=R\;,\ \ \ \ L_{2}=R_{\mu\nu\alpha\beta}R^{\mu\nu\alpha\beta}-4R_{\mu\nu}R^{\mu\nu}+R^2\;.
\end{eqnarray}
They are related to the cosmological constant, Einstein-Hilbert Lagrangian and Lanczos Lagrangian \cite{lan:1932,lan:1938}, respectively.

The variation of action with respect to the metric gives the Lovelock gravity \cite{ll:1971}
\begin{eqnarray}
\sum_{p=0}^{N}\alpha_p G^{(p)}_{\mu\nu}=\kappa T_{\mu\nu}\;.
\end{eqnarray}
$T_{\mu\nu}$ is the energy momentum tensor of matters and $\kappa$ is a constant. In four dimensional case, $\kappa=8\pi$. The Lovelock tensors are
\begin{eqnarray}
G^{(p)}_{\mu\nu}=-2^{-p-1}g_{\mu\beta}\delta_{\nu\sigma_1\sigma_2\cdot\cdot\cdot \sigma_{2p}}^{\beta\lambda_1\lambda_2\cdot\cdot\cdot \lambda_{2p}}
R_{\lambda_1\lambda_2}^{\sigma_1\sigma_2}R_{\lambda_3\lambda_4}^{\sigma_3\sigma_4}\cdot\cdot\cdot R_{\lambda_{2p-1}\lambda_{2p}}^{\sigma_{2p-1}\sigma_{2p}}\;.
\end{eqnarray}

In particular, we have \cite{bri:1997}
\begin{eqnarray}
&&G^{(0)}_{\mu\nu}=-\frac{1}{2}g_{\mu\nu}\;,\nonumber\\
&&G^{(1)}_{\mu\nu}=R_{\mu\nu}-\frac{1}{2}g_{\mu\nu}R\;.
\end{eqnarray}
The antisymmetric Kronecker delta tensor is non-vanishing only when the indices are all different. Therefore, the maximum of order $p$ must be smaller than $(n-1)/2$. For example, when $n=4$, the Lovelock gravity leads to the Einstein tensor (p=1) plus cosmological constant term (p=0). The higher orders $p\geq 2$ do not contribute the equations of motion. It is at least when $n=5$ that, the higher order $p=2$ Lovelock tensor plays a part in the equations of motion.

Therefore, in 4-dimensions, the static and spherically symmetric black hole is the Schwarzschild-de Sitter (or anti-de Sitter) black hole which is singular. The equation of motion describing the evolution of Friedmann-Robertson-Walker universe is the standard Friedmann equation which reveals a Big-Bang singularity. However, it is not the case in the regularized Lovelock gravity theory. Actually, there is a large number of solutions for nonsingular black holes and nonsingular universes. In the next sections, we will seek for these solutions.

\section{4-dimensional nonsingular black holes}
In this section, we shall show we are able to construct a large number of nonsingular black holes in the regularized Lovelock gravity theory. To this end, let's start from the black hole solution in the Lovelock gravity in $n$ dimensional spacetime which is given in \cite{Boulware:1985wk,Wheeler,Wiltshire:1985us,Cai:2001dz}
\begin{equation}\label{Lmetric}
ds^2=-f(r)dt^2+\frac{1}{f(r)}dr^2 + r^2d\Omega_{n-2}^2\;,
\end{equation}
with
\begin{equation}
f(r)=1-r^2\psi(r)\;.
\end{equation}
Here $\psi(r)$ has the dimension of inverse square of length. It is determined by solving for the real roots of the following polynomial equation
\begin{equation}\label{solu-n}
\sum_{p=0}^{N} c_p\psi^p=\frac{16\pi M}{\left(n-2\right)\Omega_{n}r^{n-1}}\;,
\end{equation}
where $\Omega_{n}=2\pi^{(n-2)/2}/\Gamma[(n-2)/2]$ is the volume of an $(n-2)$-dimensional unit sphere and the coupling constants $c_p$ are defined in terms of those appearing in Lagrangian
\begin{equation}
c_0=\frac{\alpha_0}{\alpha_1}\frac{1}{\left(n-1\right)\left(n-2\right)}\;,\ \ c_1=1\;,\ \  \ c_p=\frac{\alpha_p}{\alpha_1} \prod _{k=3}^{2p}\left(n-k\right)\;,\ \  \textrm{for} \ \ p>1\;.
\end{equation}
$M$ is the mass of black hole.

The product of factors $(n-k)$ demonstrates the fact that the $p$-th order of Lovelock tensor would not affect the field equations when $n\leq 2p$. In particular, when $n=4$, we have \cite{Boulware:1985wk,Wheeler,Wiltshire:1985us,Cai:2001dz}
\begin{eqnarray}\label{Ppsi}
c_0&=&\frac{\alpha_0}{\alpha_1}\frac{1}{\left(4-1\right)\left(4-2\right)}\;,\nonumber\\
c_1&=&1\;,\nonumber\\\
c_2&=&\frac{\alpha_2}{\alpha_1} \left(4-3\right)\left(4-4\right)\;,\nonumber\\
c_3&=&\frac{\alpha_3}{\alpha_1}\left(4-3\right)\left(4-4\right)\left(4-5\right)\left(4-6\right)\;,\nonumber\\
c_4&=&\frac{\alpha_4}{\alpha_1} \left(4-3\right)\left(4-4\right)\left(4-5\right)\left(4-6\right)\left(4-7\right)\left(4-8\right)\;,\nonumber\\
&&\cdot\cdot\cdot\cdot\cdot\cdot
\end{eqnarray}
It is apparent, if $\alpha_p$ is not regularized, the Lovelock gravity would lead us to the Schwarzschild-de Sitter solution in 4-dimensions since all $c_p$ with $p>1$ are vanishing. Following Ref.~\cite{glavan:2020,CasalinoCRV},  let's make a regularization

\begin{eqnarray}\label{Ppsi}
\alpha_p\rightarrow{\alpha_p}\left[\prod_{k=3}^{2p}\left(n-k\right)\right]^{-1}\;,\ \ \ \ \textrm{for} \ \ p>1\;,
\end{eqnarray}
Then the regularized Lovelock gravity in n-dimensions is described by
\begin{eqnarray}
{S}=\int d^n x\sqrt{-g}\left\{\alpha_0+\alpha_1 R+\sum_{p=2}\alpha_p {\left[\prod_{k=3}^{2p}\left(n-k\right)\right]^{-1}} L_{p}+L_m\right\}\;.
\end{eqnarray}
We immediately conclude that the 4-dimensional black hole solution in the regularized Lovelock gravity takes  the form
\begin{equation}\label{solu-n}
\sum_{p=0}^{N} c_p\psi^p=\frac{2 M}{r^{3}}\;.
\end{equation}
But the dimensional constants $c_p$ for $p>1$ are now non-vanishing. They are,
\begin{equation}\label{const}
c_0=\frac{\alpha_0}{6\alpha_1}\;,\ \ c_1=1\;,\ \  \ c_p=\frac{\alpha_p}{\alpha_1}\left[\prod_{k=3}^{2p}\left(n-k\right)\right]^{-1} \prod _{k=3}^{2p}\left(n-k\right)=\frac{\alpha_p}{\alpha_1}\;,\ \  \textrm{for} \ \ p>1\;.
\end{equation}
In principle, we can let
\begin{equation}
N=+\infty\;,
\end{equation}
because $N$ can be arbitrary large.
By defining
\begin{equation}\label{MEq}
P\left(\psi\right)\equiv\sum_{p=0}^{+\infty} c_p\psi^p\,,
\end{equation}
we can get any form of $P(\psi)$ we want provided that the coupling constants $c_p$ are properly chosen. We emphasize that the idea of $N=+\infty$ is firstly proposed by Kunstatter,  Maeda and Taves \cite{maeda:2006} where \emph{the regularized Lovelock gravity} is called \emph{the designer Lovelock gravity}.

In all, the four dimensional, static spherically symmetric and vacuum solution of regularized Lovelock gravity is
\begin{equation}
P\left(\psi\right)=\frac{2M}{r^3}\;.
\end{equation}
$P(\psi)$ is given by Eq.~(\ref{MEq}). The coupling constants $c_p$ are given by Eq.~(\ref{const}).
In the next subsections, we shall present several nonsingular black holes in the regularized Lovelock gravity.

\subsection{NBH-1 (nonsingular black hole-1)}
Let's consider
\begin{equation}\label{power}
P\left(\psi\right)=-\frac{\lambda}{3}+\frac{1}{\beta}\left[\left(1+\frac{\beta\psi}{\gamma}\right)^{\gamma}-1\right]=-\frac{\lambda}{3}+\psi+\frac{\beta}{2\gamma}\left(\gamma-1\right)\psi^2
+\frac{\beta^2}{6\gamma^2}\left(\gamma-1\right)
\left(\gamma-2\right)\psi^3+\cdot\cdot\cdot\,,
\end{equation}
where the constant $\gamma$ is dimensionless while $\beta$ has the dimension of square of length, and $\lambda$ is the cosmological constant. Substituting it into Eq.~(19), we obtain the expression of $\psi$. Then substituting $\psi$ into Eq.~(9), we obtain the expression of $f$ for the 4-dimensional black hole as follows
\begin{equation}\label{Lmetric}
f=1-\frac{\gamma r^2}{\beta}\left[\left(1+\frac{2\beta M}{r^3}+\frac{\beta\lambda}{3}\right)^{\frac{1}{\gamma}}-1\right]\;.
\end{equation}
Expanding it in the series of $\beta$, we obtain the Schwarzschild-(Anti)de Sitter solution by putting $\beta=0$. When $\gamma>0,\ \lambda=0$ and for sufficient large $r$, we obtain a Minkowski spacetime. When $\gamma>0,\ \lambda=0$ and for sufficient small $r$, we obtain
\begin{equation}\label{Lmetric}
f=1-\frac{\gamma}{\beta}\left(2\beta M\right)^{\frac{1}{\gamma}}r^{2-\frac{3}{\gamma}}\;.
\end{equation}
Therefore, there is physical singularity when $0<\gamma<3/2$. When $\gamma>3/2$, there is no physical singularity.
{In all, when $\gamma>3/2$ and $\beta>0$, the black hole is singularity-free and asymptotically de Sitter ($\lambda>0$) or anti de Sitter ($\lambda<0$)}.
On the other hand, we always obtain the Minkowski spacetime whether for $\gamma<0,\ \lambda=0$ when $r\rightarrow 0$ or for $\gamma<0,\ \lambda=0$ when $r\rightarrow +\infty$. {In all, when $\gamma<0$ and $\beta>0$, the black hole is singularity-free and asymptotically de Sitter ($\lambda>0$) or anti de Sitter ($\lambda<0$)}.

\begin{figure}[htbp]
	\centering
	\includegraphics[width=8cm,height=6cm]{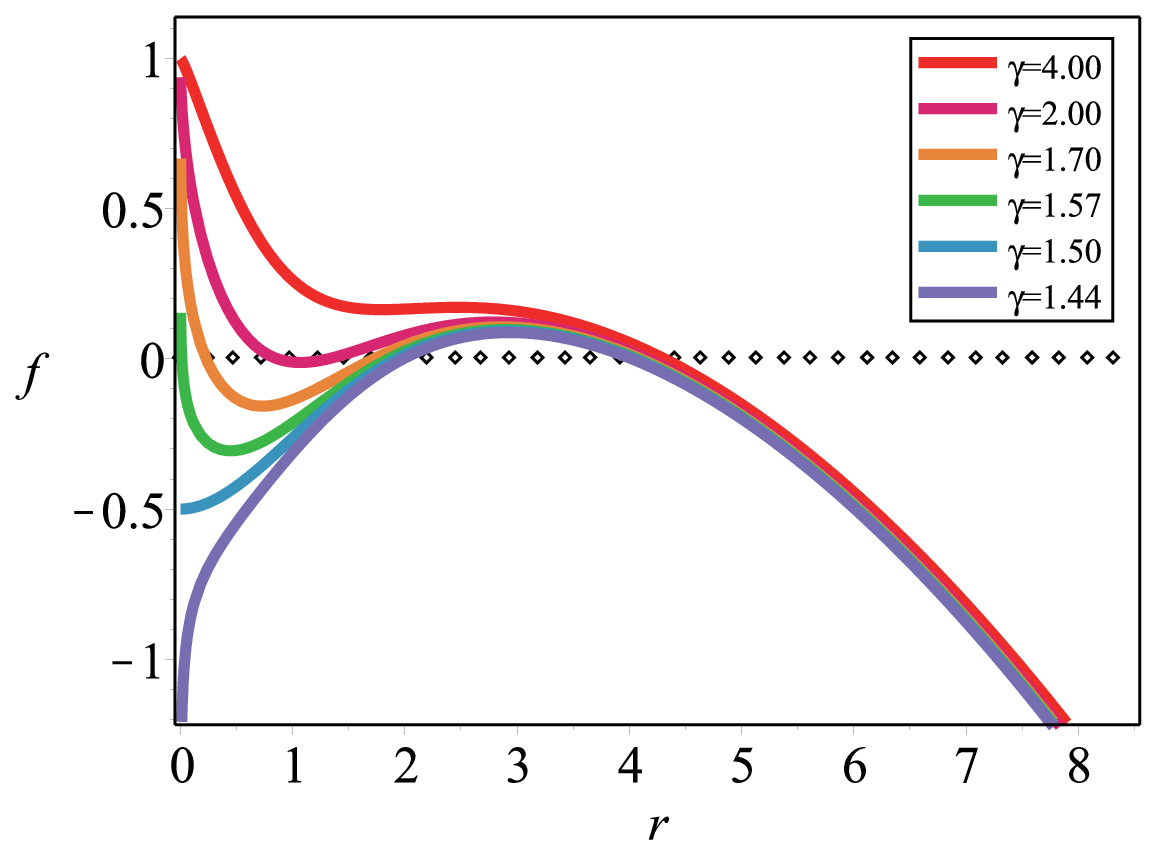}
	\caption{The evolution of $f$ with respect to $r$ for positive $\gamma$. The black hole spacetime is singularity-free when $\gamma\geq 3/2$. }\label{horizons-g}
\end{figure}

In Fig.~\ref{horizons-g}, we plot the positions of black hole horizons for positive $\gamma$. We have put $M=1,\ \beta=4,\ \lambda=0.1$ and $\gamma=4,\ 2,\ 1.7,\ 1.57,\ 1.5,\ 1.44$ from top to bottom, respectively. It shows that all the spacetimes are asymptotically de Sitter in space. When $\gamma\geq 2$, the spacetime has only one cosmic horizon.
When $3/2<\gamma<2$, the spacetime has three horizons (the outer event horizon, the inner horizon and the cosmic horizon) and no singularity. When $0<\gamma<3/2$, the spacetime has the event horizon, the cosmic horizon and a physical singularity. These plots are consistent with the previous semi-analytic analysis.

\begin{figure}[htbp]
	\centering
	\includegraphics[width=8cm,height=6cm]{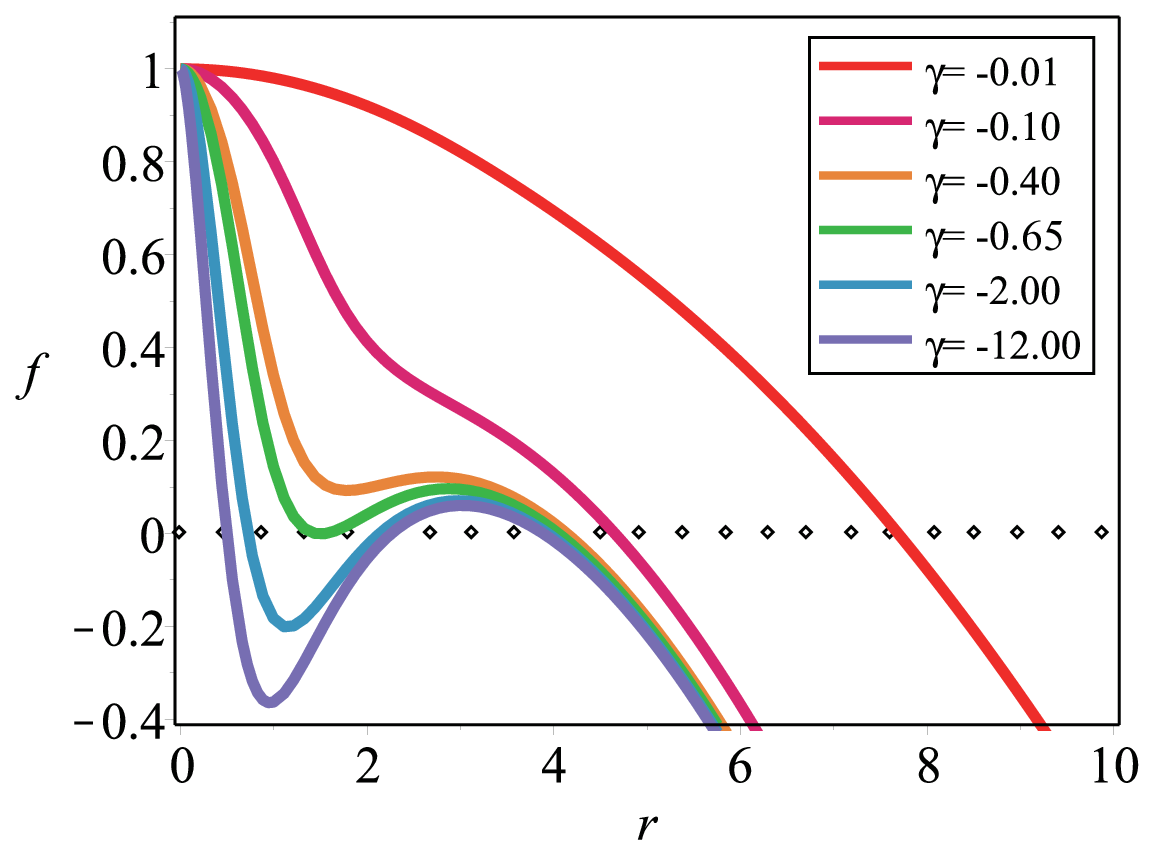}
	\caption{The evolution of $f$ with respect to $r$ for negative $\gamma$. The black hole spacetime is singularity-free. }\label{ngamma}
\end{figure}

In Fig.~\ref{ngamma}, we plot the positions of black hole horizons for negative $\gamma$. We put $M=1,\ \beta=0.5,\ \lambda=0.1$ and $\gamma=-0.01,\ -0.1,\ -0.4,\ -0.65,\ -2,\ -12$ from top to bottom, respectively. It shows that all the spacetimes are asymptotically de Sitter in space. When $-0.65<\gamma<0$, the spacetime has only one cosmic horizon and no singularity. When $\gamma<-0.65$, the spacetime has three horizons and no singularity. These plots are also consistent with the previous semi-analytic analysis.

\subsection{NBH-2 (nonsingular black hole-2)}
Now considering
\begin{equation}\label{Ppsi}
P\left(\psi\right)=-\frac{\lambda}{3}-\frac{1}{\beta}\ln\left(1-{\beta}{\psi}\right)=-\frac{\lambda}{3}+\psi+\frac{1}{2}\beta\psi^2+\frac{1}{3}\beta^2\psi^3+\frac{1}{4}\beta^3\psi^4+\cdot\cdot\cdot\,,
\end{equation}
then we obtain $f$ for a nonsingular black hole
\begin{equation}\label{Lmetric}
f=1-\frac{r^2}{\beta} \left(1-e^{-\frac{2\beta M}{r^3}-\frac{\beta\lambda}{3}}\right)\;.
\end{equation}
We assume $\beta>0$. For very large $r$, it reduces to the de Sitter ($\lambda>0$) or anti-de Sitter ($\lambda<0$) solution. When $r\rightarrow 0$, we get a de Sitter core. The computations of the Ricci scalar and the Riemann tensor reveal they are indeed not divergent ar $r=0$. Therefore, this is a nonsingular black hole solution.

\begin{figure}[htbp]
	\centering
	\includegraphics[width=8cm,height=6cm]{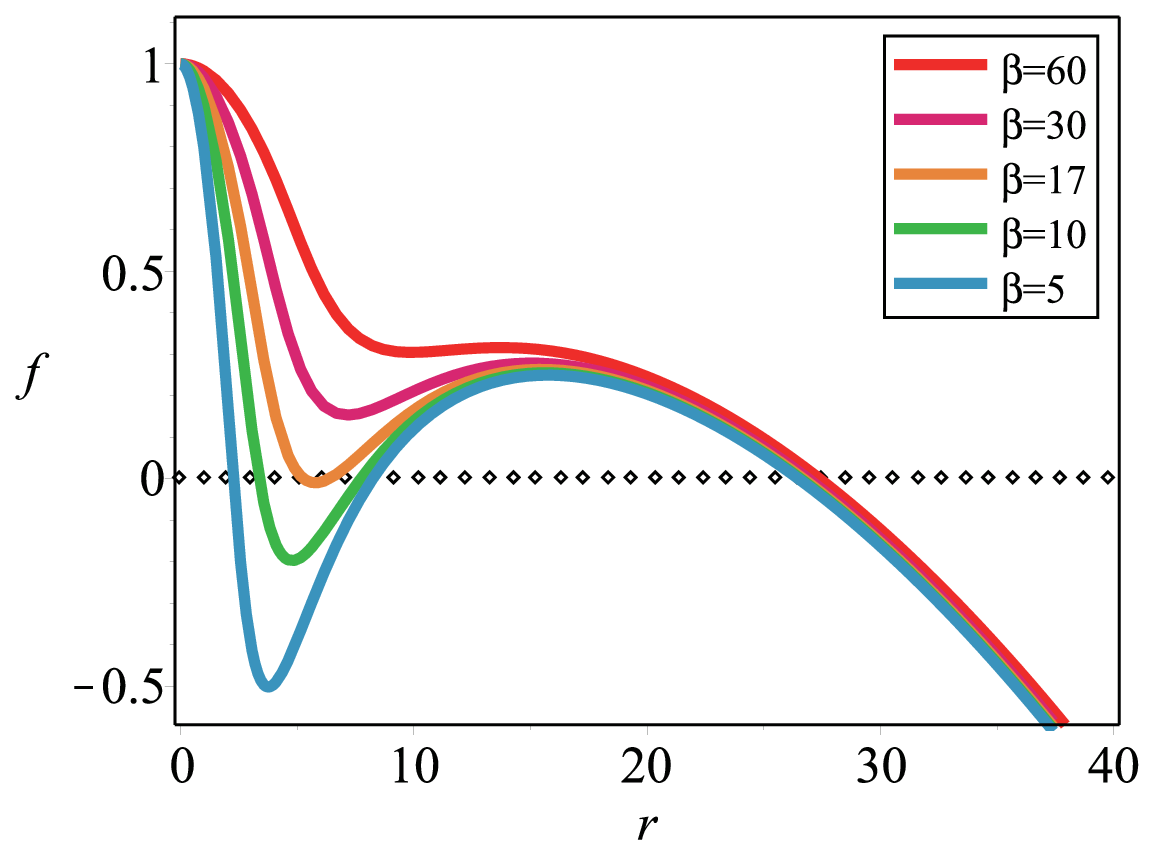}
	\caption{The evolution of $f$ with respect to $r$. The black hole has no singularity. }\label{solu-2}
\end{figure}

In Fig.~\ref{solu-2}, we plot the positions of black hole horizons. We put $M=4,\ \lambda=0.003$ and $\beta=60,\ 30,\ 17,\ 10,\ 5$ from top to bottom, respectively. It shows that all the spacetimes are asymptotically de Sitter in space. When $\beta\geq 17$, the spacetime has only one cosmic horizon and no singularity.
When $\beta<17$, the spacetime has three horizons and is singularity-free.

We have considered two nonsingular black hole solutions above. Actually, we have many forms of $P(\psi)$ for non-singular black holes such as
\begin{equation}\label{functions}
P\left(\psi\right)=-\frac{\lambda}{3}-\frac{1}{\beta}\left(1-e^{\beta\psi}\right)\;, \ \ \ \ -\frac{\lambda}{3}+\frac{1}{\beta}\sinh\left(\beta\psi\right)\;, \ \ \ \ -\frac{\lambda}{3}-\frac{1}{\beta}\left(1-\cosh{\sqrt{2\beta\psi}}\right)\;,\ \ \ \ -\frac{\lambda}{3}+\frac{\psi}{1-\beta\psi^2}\;, \ \ \ -\frac{\lambda}{3}+\frac{\psi}{1-\beta\psi}\;.
\end{equation}
These  black holes are all asymptotically de Sitter or anti-de Sitter and singularity-free. In particular, the last one, $P(\psi)=-\frac{\lambda}{3}+\frac{\psi}{1-\beta\psi}$ gives
\begin{equation}\label{Lmetric}
f=1-\frac{r^2\left(\lambda r^3+6M\right)}{3r^3+\lambda\beta r^3+6\beta M}\;.
\end{equation}
When $\lambda=0$, it is exactly the well-known Hayward nonsingular black hole \cite{hayward:2006}. It should be noted that Kunstatter et al also found the Hayward black hole from a Lovelock-like theory \cite{maeda:2006} . On the other hand, Colleaux constructed the one-parameter generalization of Hayward non-singular black hole and a non-singular universe \cite{coll:2019}.

\subsection{NBH-3 (nonsingular black hole-3)}
Now let's consider another interesting case for nonsingular black hole solution
\begin{equation}\label{Ppsi}
P\left(\psi\right)=-\frac{\lambda}{3}+\frac{1}{\beta}\arcsin\left(\beta\psi\right)=-\frac{\lambda}{3}+\psi+\frac{1}{6}\beta^2\psi^3+\frac{3}{40}\beta^4\psi^5+\cdot\cdot\cdot\,.
\end{equation}
We obtain the expression of $f$ from Eq.~(19) and Eq.~(9) for nonsingular black hole
\begin{equation}\label{Lmetric}
f=1-\frac{r^2 \sin{\left(\frac{2\beta M}{r^3}+\frac{\lambda\beta}{3}\right)}}{\beta}\;.
\end{equation}
We assume $\beta>0$. For very large $r$ and $|\lambda\beta|\ll 1$, it reduces to the de Sitter or anti-de Sitter solution. When $r\rightarrow 0$, we get the Minkowski spacetime. We find the Ricci scalar and the Riemann tensor are not divergent ar $r=0$. Therefore, this is a nonsingular black hole solution.

\begin{figure}[htbp]
	\centering
	\includegraphics[width=8cm,height=6cm]{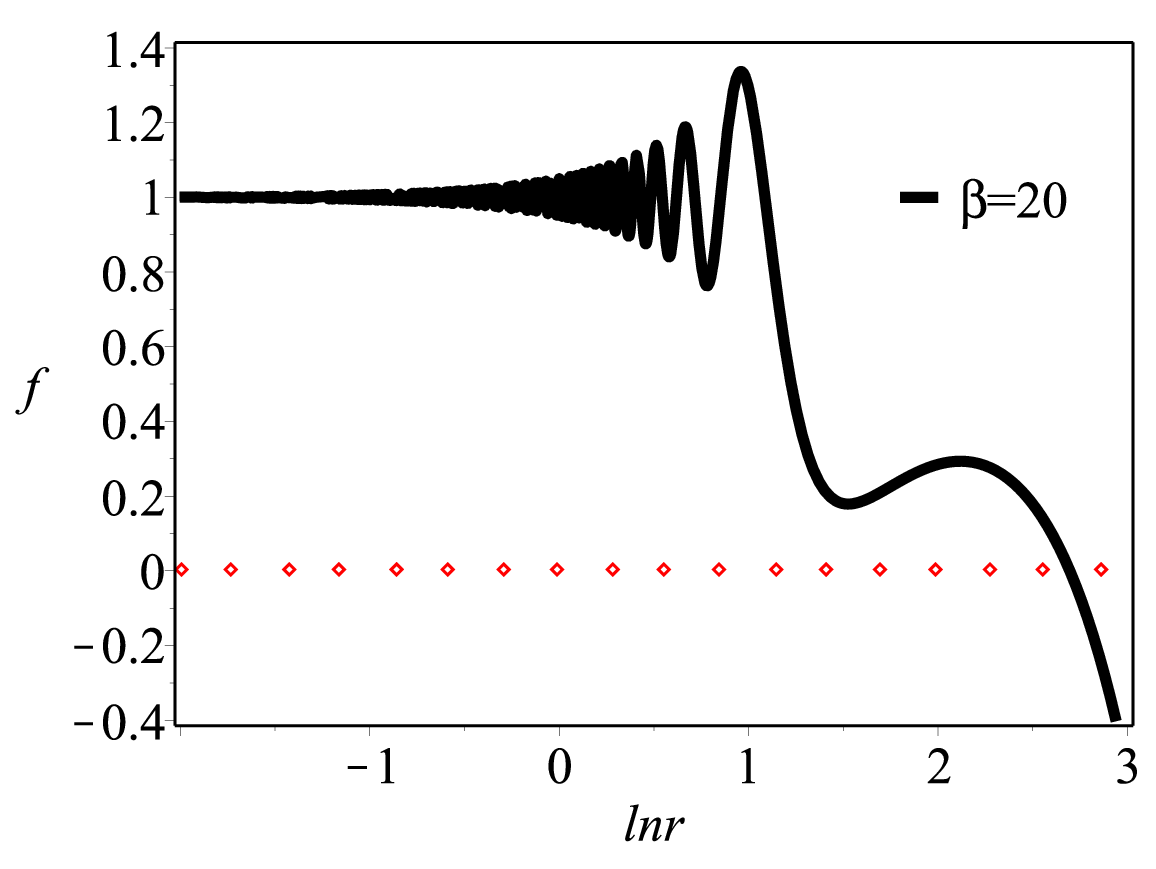}
	\caption{The evolution of $f$ with respect to $\ln{r}$ when $M=2,\ \lambda=0.01,\ \beta=20$. In this case, the spacetime has only one cosmic horizon and no singularity}\label{solu-3a}
\end{figure}

\begin{figure}[htbp]
	\centering
	\includegraphics[width=8cm,height=6cm]{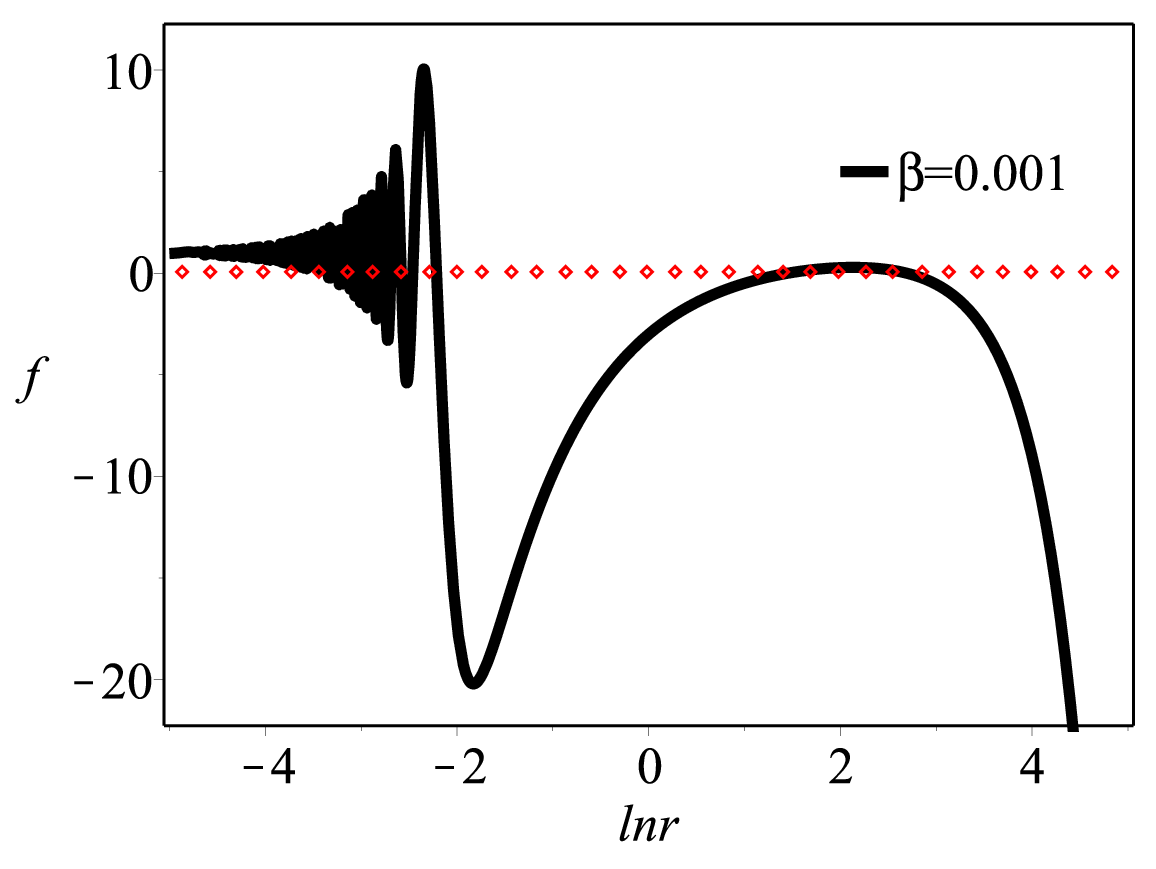}
	\caption{The evolution of $f$ with respect to $\ln{r}$ when $M=2,\lambda=0.01,\ \beta=0.001$. The spacetime is singularity-free and has many horizons.}\label{solu-3b}
\end{figure}

In Fig.~\ref{solu-3a}, we plot the positions of black hole horizons with $M=2,\ \lambda=0.01$ and $\beta=20$. It shows that, in this case, there is only one cosmic horizon in this spacetime.
In Fig.~\ref{solu-3b}, we plot the positions of black hole horizons with $M=2,\ \lambda=0.01$ and $\beta=0.001$. It shows that there are many horizons in this spacetime.
Detailed analysis reveals that, when $0<\beta<9.5$ for $M=2$, there are many horizons in this spacetime, one cosmic horizon and large number of black hole horizons. The smaller $\beta$, the more black hole horizons. We note that the concept of \emph{singular and nonsingular multi-horizon black
holes }in general relativity and modified gravity with nonlinear electrodynamics is already studied in Ref.~\cite{gao:2018}. But their starting points are different from this paper. To gain more properties of these nonsingular black holes, in the next sections, we shall calculate their quasinormal modes and geodesic motions.

\section{Quasinormal modes}

When a black hole is perturbed, it undergoes a long period of damping oscillations. The process of this damping oscillations is dominated by the so-called quasinormal modes. It is this quasinormal modes that dominates the contribution to gravitational waves. In this section we study the quasinormal modes generated by scalar perturbations of the nonsingular black holes. To this end, we start from the scalar perturbation equation
\begin{eqnarray}
\nabla^2\Psi=0\;,
\end{eqnarray}
which is the general perturbation equation for the massless scalar field in the curved spacetime.
Here $\nabla^2$ is the four dimensional Laplace operator and $\Psi$ the massless scalar field. Making the standard decomposition
\begin{eqnarray}
\Psi=e^{-i\omega t}Y_{lm}\left(\theta,\ \phi\right)\frac{\Phi\left(r\right)}{r}\;,
\end{eqnarray}
we obtain the radial perturbation equation
\begin{eqnarray}
\frac{d^2\Phi}{d r_{\ast}^2}+\left(\omega^2-V\right)=0\;,
\end{eqnarray}
where the effective potential is given by
\begin{eqnarray}
V=f\left(\frac{l\left(l+1\right)}{r^2}+\frac{f_{,r}}{r}\right)\;,
\end{eqnarray}
and $r_{\ast}$ is the tortoise coordinate defined by
\begin{eqnarray}
r_{\ast}=\int \frac{1}{f} dr\;.
\end{eqnarray}
\begin{figure}[htbp]
\centering
\subfigure[]{\label{Fig1Vf1}
\includegraphics[width=3.9cm,height=2.925cm]{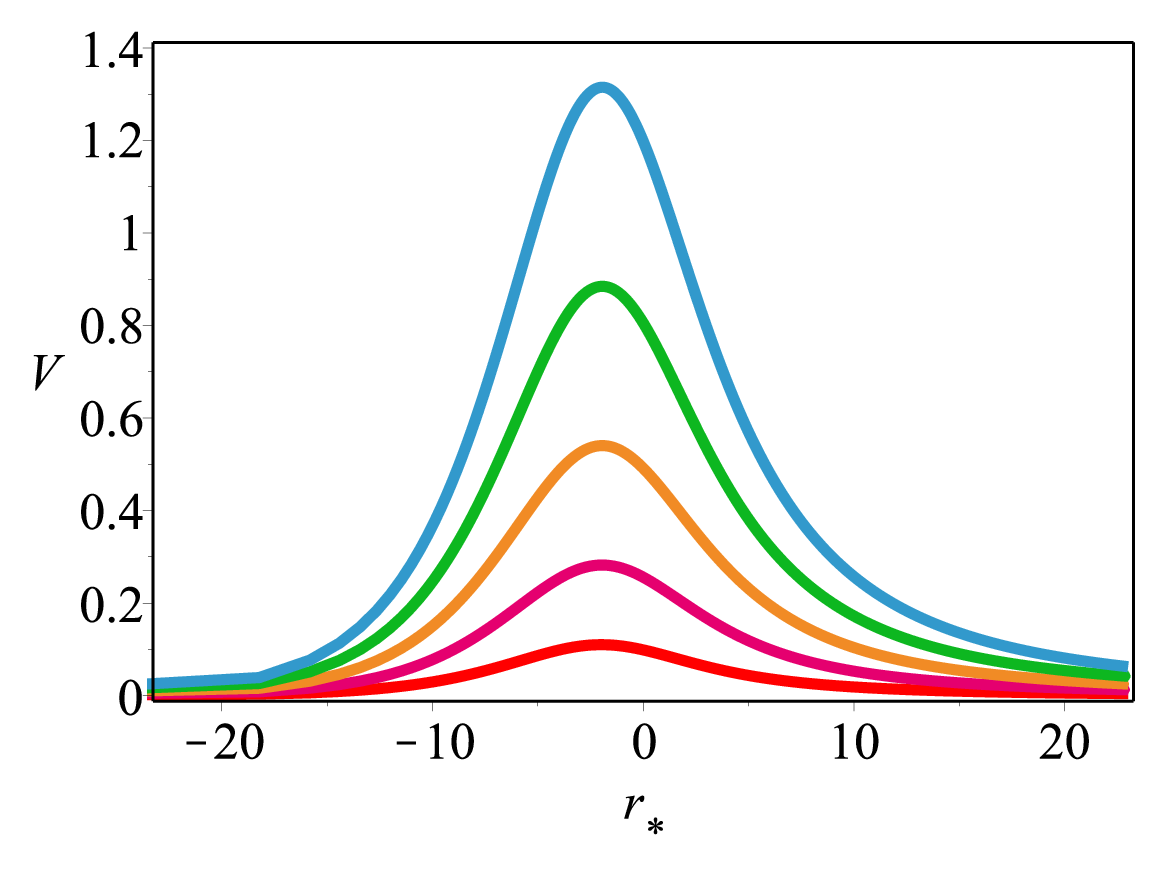}}\subfigure[]{\label{Fig1Vf2}
\includegraphics[width=3.9cm,height=2.925cm]{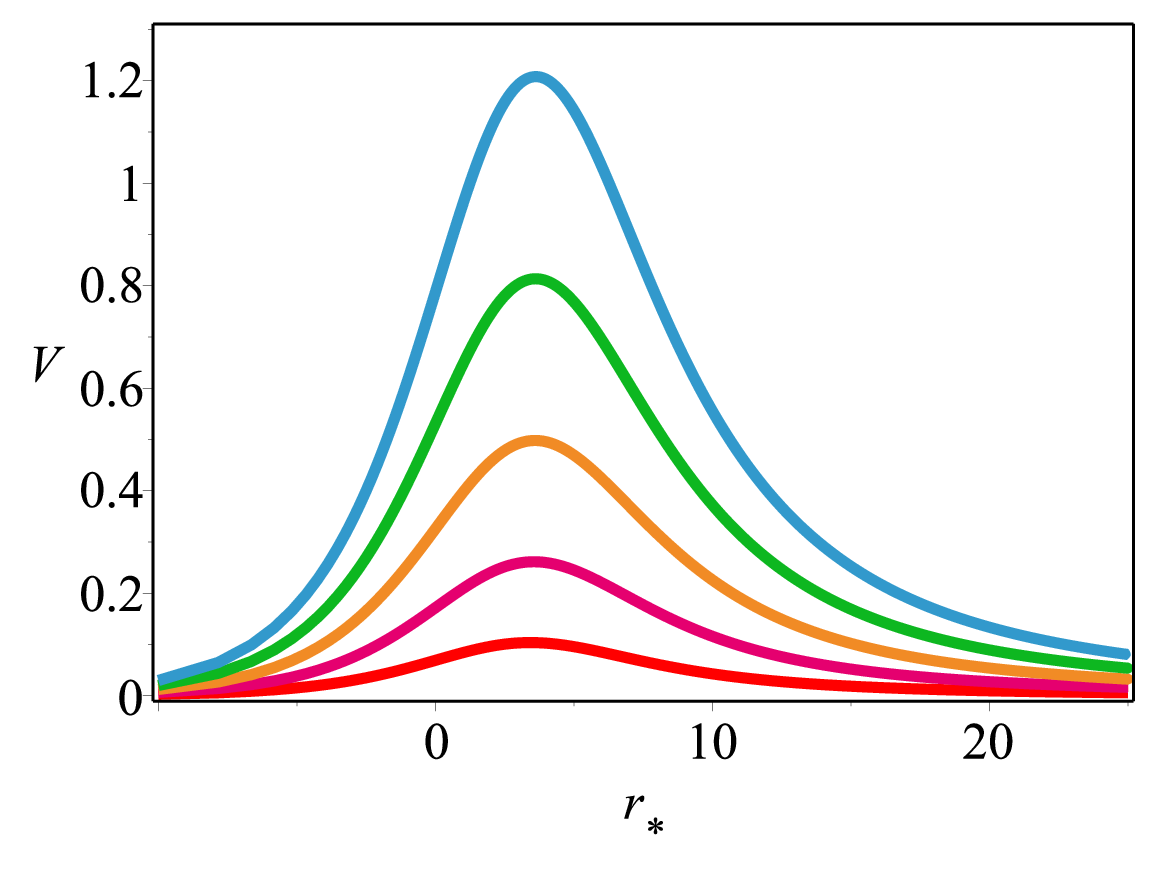}}
\subfigure[]{   \label{Fig2Vf}
\includegraphics[width=3.9cm,height=2.925cm]{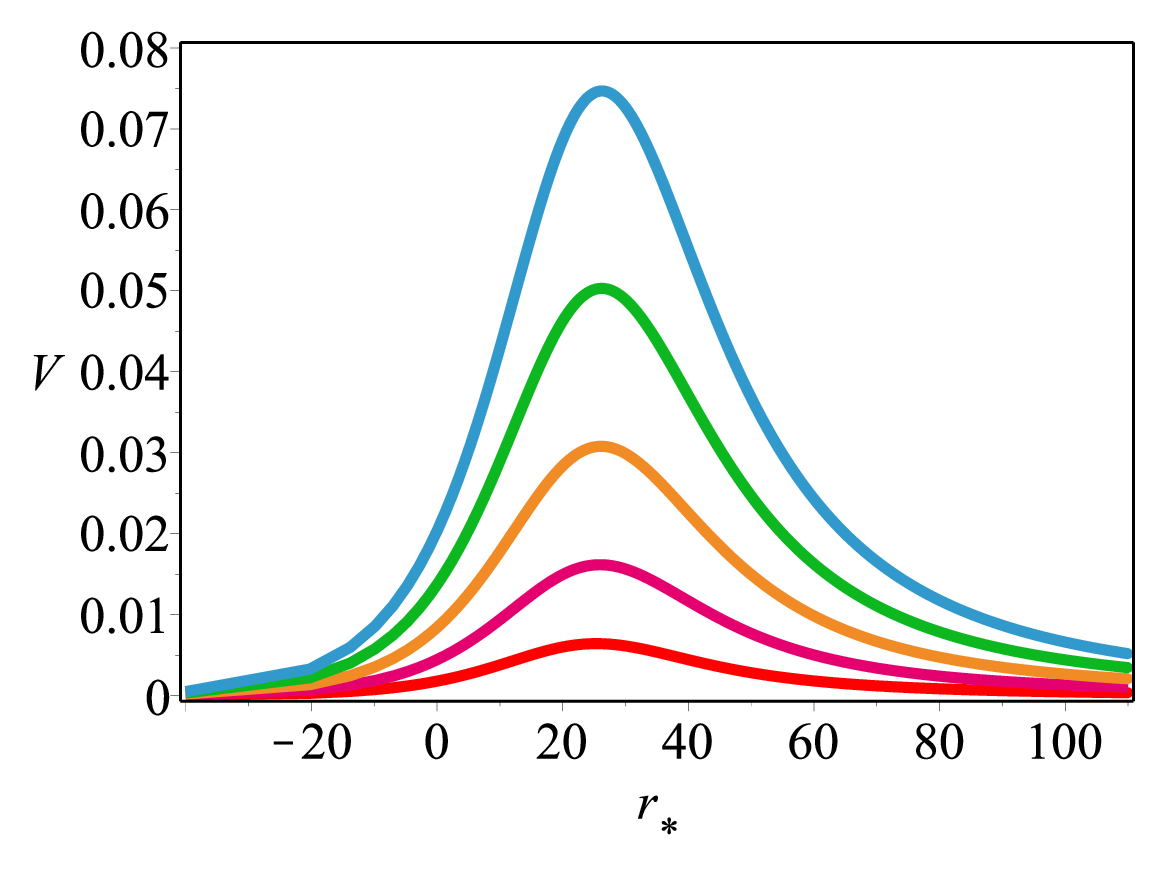}}\subfigure[]{\label{Fig3Vf}
\includegraphics[width=3.9cm,height=2.925cm]{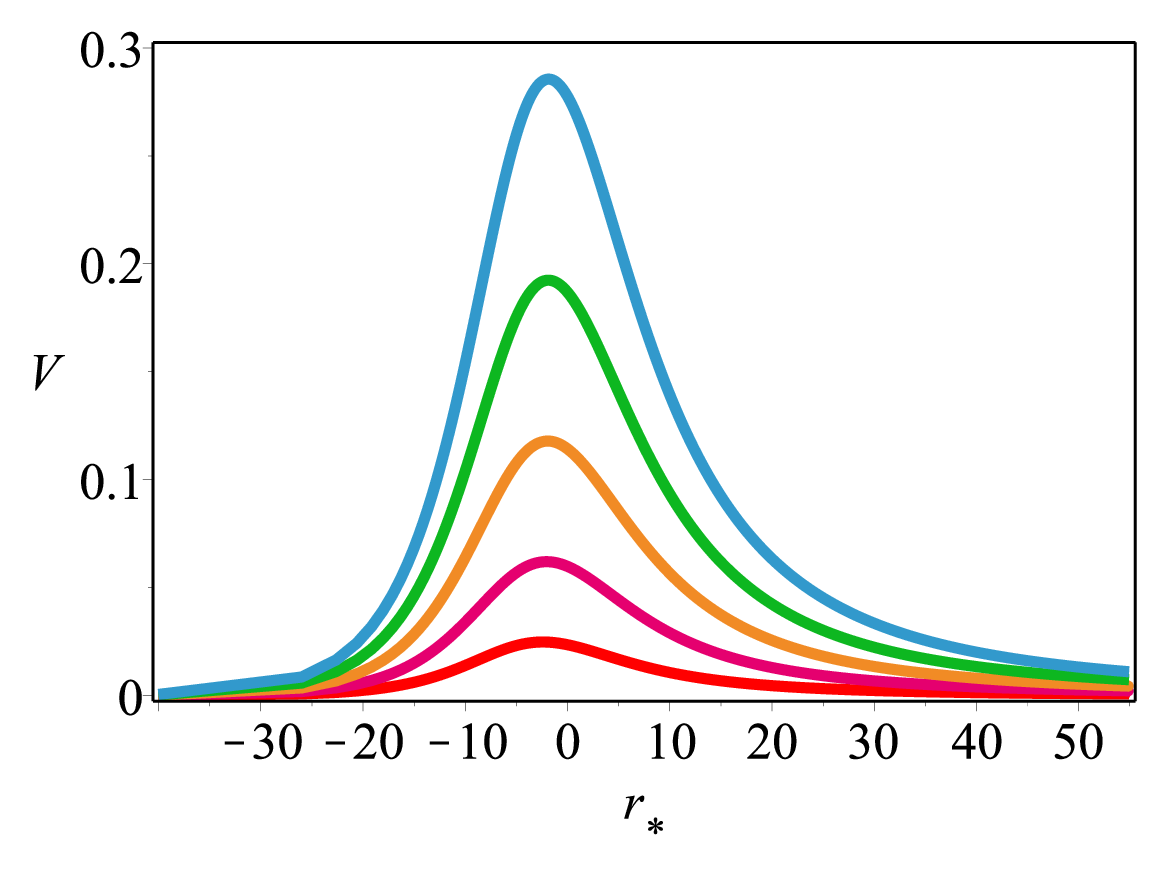}}\\
\caption{The effective potential $V(r_{\ast})$ as a function of the tortoise coordinate $r_\ast$ when $\lambda=0$ and $l=5,4,3,2,1$, from up to down. (a) $M=1, \beta=4,\gamma=1.7$ for NBH-1. (b) $M=1, \beta=0.5,\gamma=-2$ for nonsingular NBH-1. (c) $M=4, \beta=10$, for nonsingular NBH-2. (d) $M=2, \beta=7$ for NB-3.}
\end{figure}

\begin{figure}[htbp]
\centering
\subfigure[]{\label{Fig1Vd1}
\includegraphics[width=3.9cm,height=2.925cm]{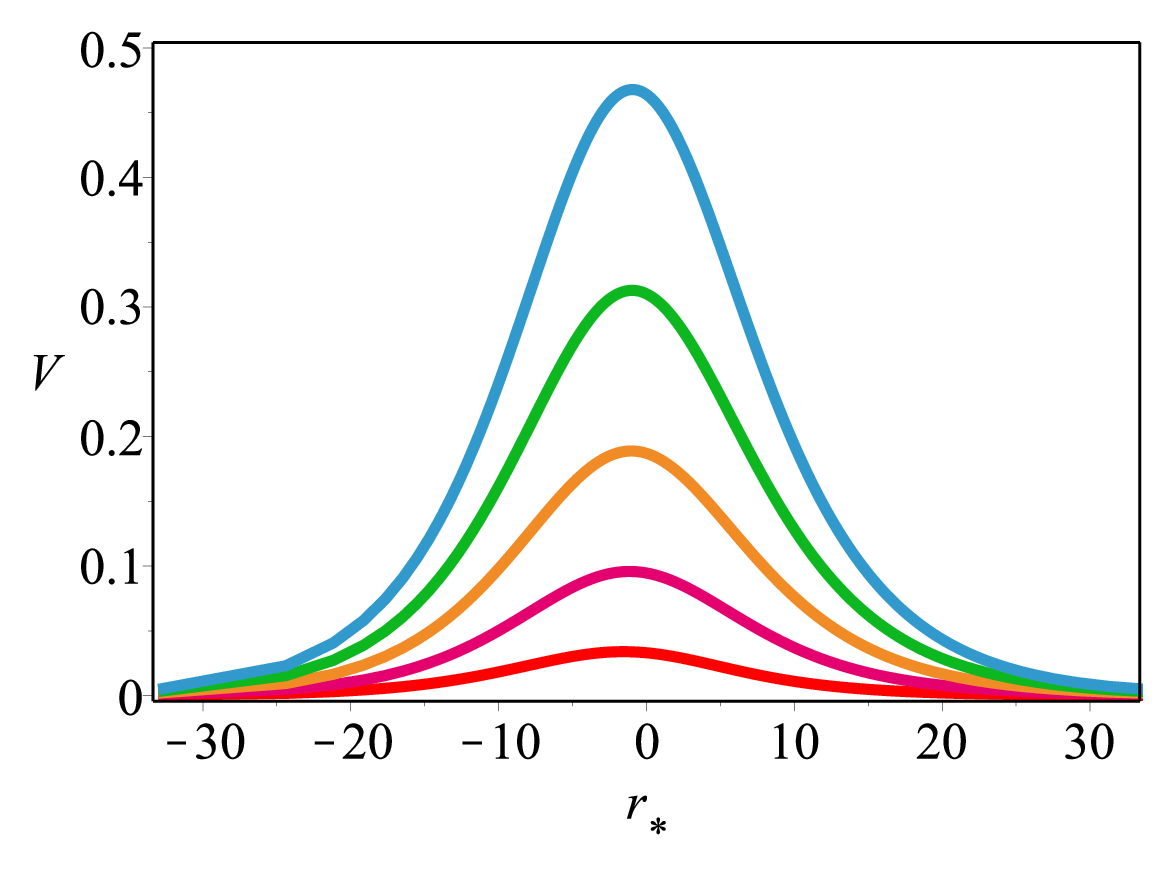}}\subfigure[]{\label{Fig1Vd2}
\includegraphics[width=3.9cm,height=2.925cm]{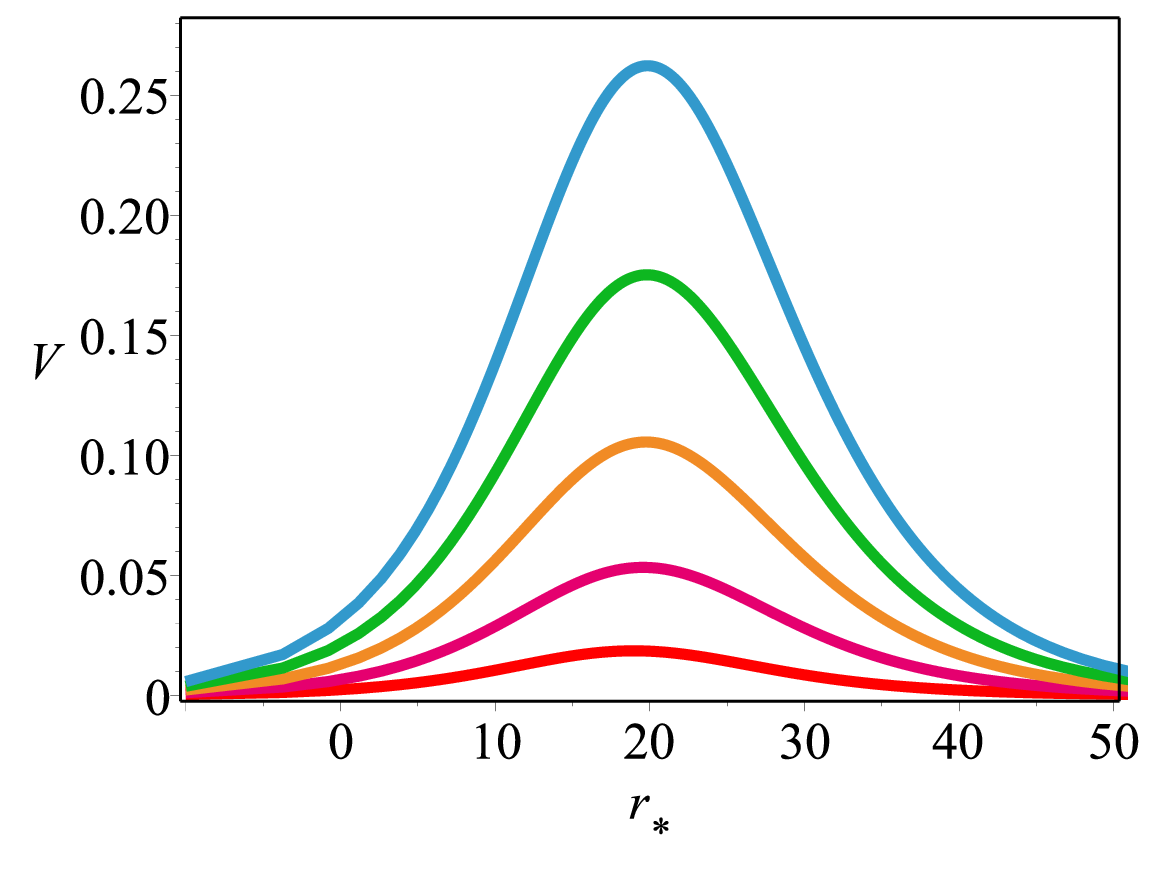}}
\subfigure[]{   \label{Fig2Vd}
\includegraphics[width=3.9cm,height=2.925cm]{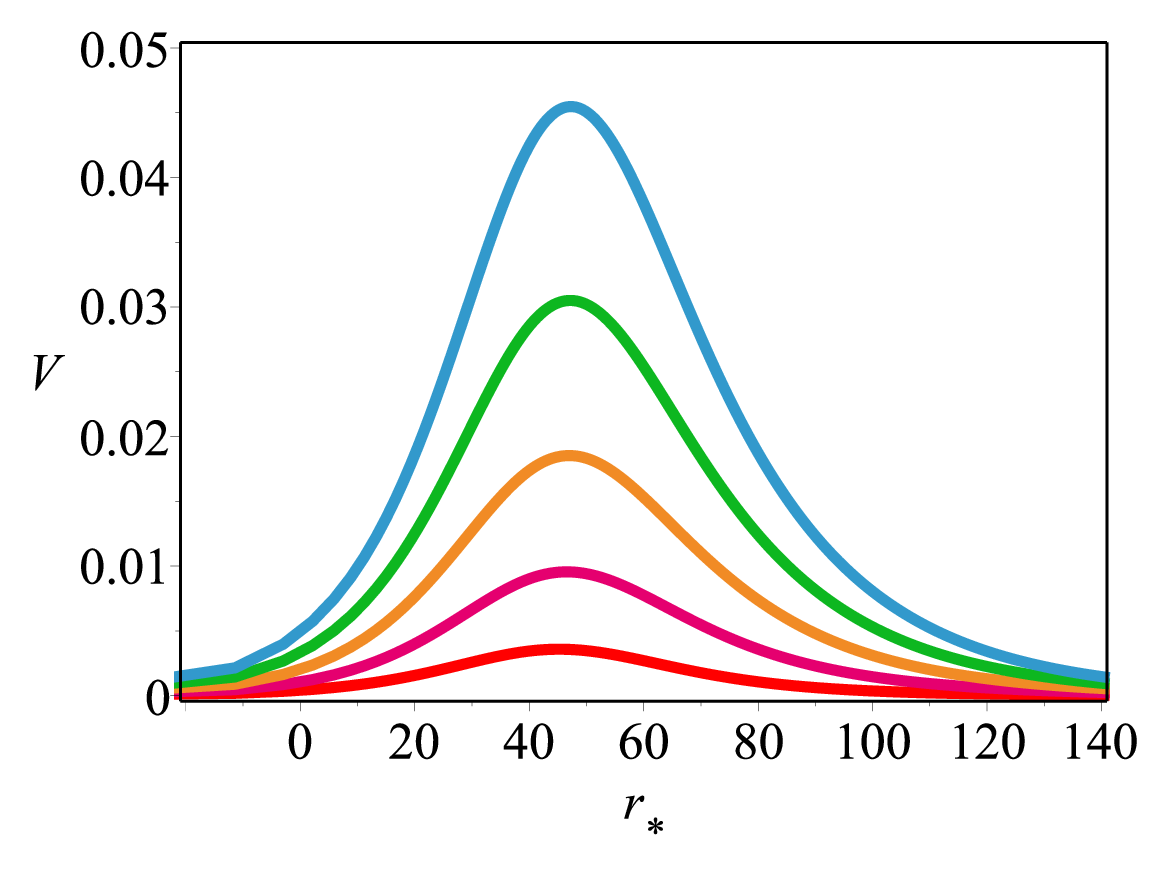}}\subfigure[]{\label{Fig3Vd}
\includegraphics[width=3.9cm,height=2.925cm]{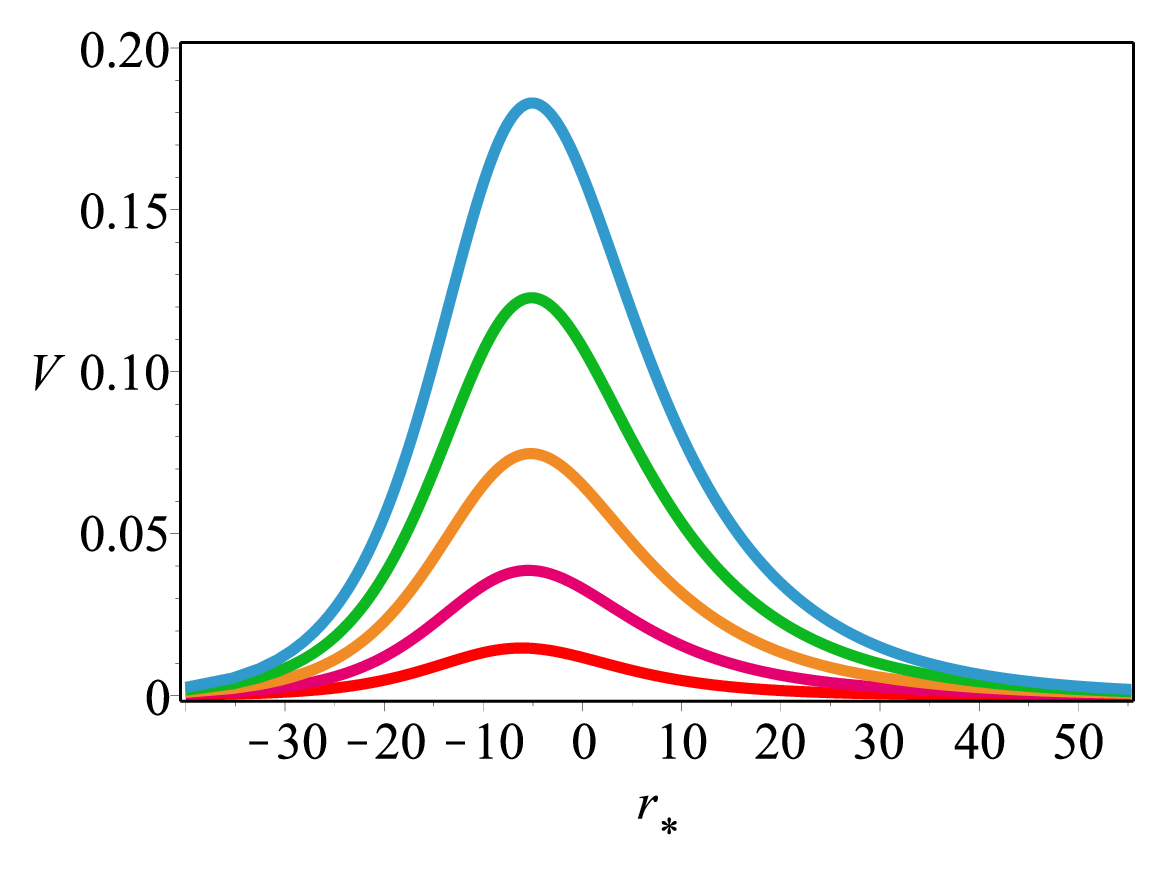}}\\
\caption{The effective potential $V(r_{\ast})$ as a function of the tortoise coordinate $r_\ast$ when the cosmological constant is non-vanishing. The plots correspond to  $l=5,4,3,2,1$ from up to down. (a) $M=1, \beta=4, \gamma=1.7, \lambda=0.1$ for NBH-1. (b) $M=1, \beta=0.5, \gamma=-2, \lambda=0.1$ for NBH-1. (c) $M=4, \beta=10, \lambda=0.003$ for NBH-2. (d) $M=2, \beta=7, \lambda=0.01$ for NBH-3.}
\end{figure}
The effective potential $V$ as a function of the tortoise coordinate $r_\ast$ of three nonsingular black holes can be seen in Fig.~\ref{Fig1Vf1},\ \ref{Fig1Vf2}, \ref{Fig2Vf},\ \ref{Fig3Vf},\ \ref{Fig1Vd1},\ \ref{Fig1Vd2},\ \ref{Fig2Vd},\ \ref{Fig3Vd}. Comparing Fig.~\ref{Fig1Vf1},\ \ref{Fig1Vf2},\ \ref{Fig2Vf},\ \ref{Fig3Vf} (asymptotically flat black holes) with Fig.~\ref{Fig1Vd1},\ \ref{Fig1Vd2},\ \ref{Fig2Vd},\ \ref{Fig3Vd} (asymptotically de Sitter black holes), we find that the presence of a positive cosmological constant decreases the height of effective potential. We also find that the height of effective potential  increases with the increasing of the multi-pole number $l$. We shall evaluate the quasinormal frequencies for the scalar perturbations by
using the third-order WKB approximation, a numerical and perhaps the most popular method, devised by Schutz, Will and Iyer \cite{will:1985,will:1987,iyer:1987}.
This method has been used extensively in evaluating quasinormal frequencies of various black holes. For an incomplete list see \cite {quasi:99} and references therein.

The quasinormal frequencies are given by
\begin{eqnarray}
\omega^2=V_0+\Lambda\sqrt{-2V_0^{''}}-i\nu\left(1+\Omega\right)\sqrt{-2V_0^{''}}\;,
\end{eqnarray}
where $\Lambda$ and $\Omega$ are
\begin{eqnarray}
\Lambda&=&\frac{1}{\sqrt{-2V_0^{''}}}\left\{\frac{V_0^{(4)}}{V_0^{''}}\left(\frac{1}{32}+\frac{1}{8}\nu^2\right)
\right.\nonumber\\&&\left.-\left(\frac{V_0^{'''}}{V_0^{''}}\right)^2\left(\frac{7}{288}+\frac{5}{24}\nu^2\right)\right\}\;,\\
\Omega&=&\frac{1}{\sqrt{-2V_0^{''}}}\left\{\frac{5}{6912}\left(\frac{V_0^{'''}}{V_0^{''}}\right)^4\left(77+188\nu^2\right)
\right.\nonumber\\&&\left.-\frac{1}{384}\left(\frac{V_0^{'''2}V_0^{(4)}}{V_0^{''3}}\right)\left(51+100\nu^2\right)
\right.\nonumber\\&&\left.+\frac{1}{2304}\left(\frac{V_0^{(4)}}{V_0^{''}}\right)^2\left(67+68\nu^2\right)\right.\nonumber\\&&\left.
+\frac{1}{288}\left(\frac{V_0^{'''}V_0^{(5)}}{V_0^{''2}}\right)\left(19+28\nu^2\right)\right.\nonumber\\&&\left.-\frac{1}{288}\left(\frac{V_0^{(6)}}{V_0^{''}}
\left(5+4\nu^2\right)\right)\right\}\;,
\end{eqnarray}
and
\begin{eqnarray}
\nu=n+\frac{1}{2}\;,\ \ \ \ V_0^{(s)}=\frac{d^sV}{dr_{\ast}^s}|_{r_{\ast}=r_p}\;,
\end{eqnarray}
$n$ is overtone number and $r_p$ corresponds to the peak of the effective
potential. It is pointed that \cite{car:04} the accuracy of the WKB method depends on the multi-pole
number $l$ and the overtone number $n$. The WKB approach is consistent with the numerical method very well provided that  $l>n$. Therefore we shall present the quasinormal frequencies of scalar perturbation for $n=0$ and $l=1,2,3,4,5$, respectively.

In order to understand the effect of cosmological constant on fundamental quasinormal frequencies of three black holes, we study two situations (with and without the cosmological constant). The fundamental quasinormal frequencies of the scalar perturbation are listed in tables I-VIII. From the tables we see with the increasing of multi-pole number $l$, the real part of the frequencies are increasing but the imaginary part of the frequencies are decreasing regardless of the presence or absence of cosmological constant. We know the real and imaginary parts describe the oscillation and damping of the modes, respectively. Therefore, we conclude that with the increasing of multi-pole number $l$, the oscillation of modes becomes faster and faster while the damping becomes slower and slower.

For NBH-1, Table-I and Table-II show that with the increasing of positive $\gamma$, the real part of the frequencies are increasing but the imaginary part are decreasing. For NBH-1, Table-III and Table-IV show that with the increasing of negative $\gamma$, both the real part and imaginary part of the frequencies are increasing for NBH-1.
For NBH-2, Table-V and Table-VI show that with the increasing of $\beta$, the real part of the frequencies are increasing but the imaginary part are decreasing.
The case of NBH-3 is subtle. For NBH-3, Table-VII tells us with the increasing of $\beta$ and for $l>1$, the real part of the frequencies are increasing but the imaginary part are decreasing. However, for $l=1$, Table-VII reveals with the increasing of $\beta$, both the real part and the imaginary part of the frequencies are decreasing. {It is different from the case for non-vanishing cosmological constant just as shown in Table-VIII. The real part of the frequencies are increasing but the imaginary part are decreasing for all $l>0$}. The reason for this point may be related to the property of multi-horizon structure for NBH-3. So in the next section, we shall pay attention to the geodesic motions of test particles in the background of NBH-3.

\begin{table}[htbp]
\begin{center}
\begin{tabular}[b]{ccccc}
 \hline \hline
 $l$&\;$\omega (\gamma=1.7)$\;&\;$\omega (\gamma=1.8)$\;&\;$\omega (\gamma=1.9)$\;&$\omega (\gamma=2.0)$\; \\ \hline
1&0.31592-0.08771I&0.31796-0.08498I&0.31968-0.08216I&0.32098-0.07924I \\
2&0.52164-0.08290I&0.52598-0.08004I&0.53008-0.07696I&0.53389-0.07358I \\
3&0.72828-0.08153I&0.73461-0.07865I&0.74076-0.07550I&0.74672-0.07200I \\
4&0.93525-0.08097I&0.94350-0.07808I&0.95161-0.07491I&0.95958-0.07136I \\
5&1.14237-0.08069I&1.15253-0.07780I&1.16255-0.07462I&1.17249-0.07104I \\
\hline \hline
\end{tabular}\label{table1f1}
\end{center}
\caption{The fundamental ($n=0$) quasinormal frequencies of NBH-1 when $M=1$,$\beta=4$,$\lambda=0.0$.}
\end{table}

\begin{table}[htbp]
\begin{center}
\begin{tabular}[b]{ccccc}
 \hline \hline
 $l$&$\omega (\gamma=1.7)$\;&\;$\omega (\gamma=1.8)$\;&\;$\omega (\gamma=1.9)$\;&\;$\omega (\gamma=2.0)$\; \\ \hline
1&0.17686-0.05250I&0.18606-0.05190I&0.19487-0.05061I&0.20330-0.04849I\\
2&0.30522-0.04930I&0.32041-0.04881I&0.33511-0.04764I&0.34957-0.04558I\\
3&0.43147-0.04845I&0.45263-0.04799I&0.47318-0.04685I&0.49351-0.04481I\\
4&0.55694-0.04811I&0.58410-0.04766I&0.61049-0.04653I&0.63667-0.04449I\\
5&0.68207-0.04794I&0.71523-0.04749I&0.74746-0.04637I&0.77947-0.04433I\\
\hline \hline
\end{tabular}\label{table1d1}
\end{center}
\caption{The fundamental ($n=0$) quasinormal frequencies of NBH-1 when $M=1$,$\beta=4$,$\lambda=0.1$.}
\end{table}

\begin{table}[htbp]
\begin{center}
\begin{tabular}[b]{ccccc}
 \hline \hline
 $l$&$\omega (\gamma=-2.0)$\;&\;$\omega (\gamma=-1.8)$\;&\;$\omega (\gamma=-1.6)$\;&\;$\omega (\gamma=-1.4)$\; \\ \hline
1&0.30227-0.09825I&0.30258-0.09790I&0.30292-0.09739I&0.30337-0.09673I\\
2&0.49900-0.09235I&0.49961-0.09196I&0.50041-0.09146I&0.50144-0.09080I\\
3&0.69690-0.09066I&0.69781-0.09027I&0.69898-0.08978I&0.70050-0.08911I\\
4&0.89510-0.08997I&0.89630-0.08959I&0.89783-0.08910I&0.89982-0.08843I\\
5&1.09346-0.08964I&1.09493-0.08926I&1.09681-0.08876I&1.09927-0.08810I\\
\hline \hline
\end{tabular}\label{table1f2}
\end{center}
\caption{The fundamental ($n=0$) quasinormal frequencies of NBH-1 when $M=1$,$\beta=0.5$,$\lambda=0.0$.}
\end{table}

\begin{table}[htbp]
\begin{center}
\begin{tabular}[b]{ccccc}
 \hline \hline
 $l$&$\omega (\gamma=-2.0)$\;&\;$\omega (\gamma=-1.8)$\;&\;$\omega (\gamma=-1.6)$\;&\;$\omega (\gamma=-1.4)$\; \\ \hline
1&0.12946-0.046107I&0.13134-0.046456I&0.13350-0.046701I&0.13638-0.04704I\\
2&0.22688-0.043015I&0.22986-0.043280I&0.23361-0.043581I&0.23847-0.04392I\\
3&0.32201-0.042247I&0.32619-0.042509I&0.33144-0.042799I&0.33825-0.04314I\\
4&0.41633-0.041936I&0.42171-0.042200I&0.42847-0.042493I&0.43722-0.04283I\\
5&0.51029-0.041792I&0.51686-0.042053I&0.52512-0.042340I&0.53582-0.04268I\\
\hline \hline
\end{tabular}\label{table1d2}
\end{center}
\caption{The fundamental ($n=0$) quasinormal frequencies of NBH-1 when $M=1$,$\beta=0.5$,$\lambda=0.1$.}
\end{table}

\begin{table}[htbp]
\begin{center}
\begin{tabular}[b]{ccccc}
 \hline \hline
 $l$&$\omega (\beta=4)$\;&\;$\omega (\beta=6)$\;&\;$\omega (\beta=8)$\;&\;$\omega (\beta=10)$\; \\ \hline
1&0.07427-0.02629I&0.07460-0.02595I&0.07493-0.02559I&0.07526-0.02520I\\
2&0.12216-0.02459I&0.12278-0.02429I&0.12342-0.02395I&0.12409-0.02357I\\
3&0.17051-0.02409I&0.17139-0.02380I&0.17231-0.02346I&0.17327-0.02309I\\
4&0.21898-0.02388I&0.22011-0.02359I&0.22130-0.02326I&0.22255-0.02289I\\
5&0.26749-0.02378I&0.26887-0.02348I&0.27033-0.02316I&0.27186-0.02279I\\
\hline \hline
\end{tabular}\label{table2f}
\end{center}
\caption{The fundamental ($n=0$) quasinormal frequencies of NBH-2 when $M=4$,$\lambda=0.0$.}
\end{table}

\begin{table}[htbp]
\begin{center}
\begin{tabular}[b]{ccccc}
 \hline \hline
 $l$&$\omega (\beta=4)$\;&\;$\omega (\beta=6)$\;&\;$\omega (\beta=8)$\;&\;$\omega (\beta=10)$\; \\ \hline
1&0.05420-0.020716I&0.05495-0.02056I&0.05573-0.020374I&0.05653-0.02014I\\
2&0.09178-0.019015I&0.09302-0.01890I&0.09431-0.018753I&0.09566-0.01856I\\
3&0.12924-0.018527I&0.13096-0.01842I&0.13275-0.018286I&0.13463-0.01810I\\
4&0.16659-0.018326I&0.16878-0.01823I&0.17108-0.018094I&0.17349-0.01792I\\
5&0.20388-0.018225I&0.20655-0.01813I&0.20935-0.017996I&0.21229-0.01782I\\
\hline \hline
\end{tabular}\label{table2d}
\end{center}
\caption{The fundamental ($n=0$) quasinormal frequencies of NBH-2 when $M=4$,$\lambda=0.003$.}
\end{table}

\begin{table}[htbp]
\begin{center}
\begin{tabular}[b]{ccccc}
 \hline \hline
 $l$&$\omega (\beta=7.0)$\;&\;$\omega (\beta=7.5)$\;&\;$\omega (\beta=8.0)$\;&\;$\omega (\beta=8.5)$\; \\ \hline
1&0.14662-0.05296I&0.14651-0.05286I&0.14639-0.05275I&0.14625-0.05263I\\
2&0.24217-0.04916I&0.24218-0.04900I&0.24219-0.04883I&0.24220-0.04864I\\
3&0.33838-0.04806I&0.33847-0.04789I&0.33856-0.04770I&0.33865-0.04749I\\
4&0.43474-0.04761I&0.43489-0.04743I&0.43504-0.04723I&0.43520-0.04701I\\
5&0.53116-0.04738I&0.53135-0.04720I&0.53156-0.04699I&0.53179-0.04677I\\
\hline \hline
\end{tabular}\label{table3f}
\end{center}
\caption{The fundamental ($n=0$) quasinormal frequencies of NBH-3 when $M=2$,$\lambda=0.0$.}
\end{table}


\begin{table}[htbp]
\begin{center}
\begin{tabular}[b]{ccccc}
 \hline \hline
 $l$&$\omega (\beta=7.0)$\;&\;$\omega (\beta=7.5)$\;&\;$\omega (\beta=8.0)$\;&\;$\omega (\beta=8.5)$\; \\ \hline
1&0.11348-0.04335I&0.11352-0.04320I&0.11355-0.04304I&0.11358-0.04286I\\
2&0.19178-0.03966I&0.19197-0.03949I&0.19216-0.03931I&0.19237-0.03911I\\
3&0.26988-0.03860I&0.27017-0.03843I&0.27048-0.03825I&0.27080-0.03804I\\
4&0.34777-0.03816I&0.34817-0.03800I&0.34857-0.03781I&0.34901-0.03761I\\
5&0.42555-0.03794I&0.42603-0.03778I&0.42654-0.03759I&0.42710-0.03739I\\
\hline \hline
\end{tabular}\label{table3d}
\end{center}
\caption{The fundamental ($n=0$) quasinormal frequencies of NBH-3 when $M=2$,$\lambda=0.01$.}
\end{table}

\section{geodesic motions of test particles}
There are at most three horizons in the spacetime of NBH-1 and NBH-2. These horizons are cosmic horizon, black hole event horizon and black hole inner horizon. Except for the non-singularity of the center, the two spacetimes are similar to the well-known Reissner-Nordstrom-de Sitter solution. However, to our knowledge, NBH-3 is a completely new solution. It is not only nonsingular but also multi-horizonal. Therefore, in this section, we shall investigate the geodesics of NBH-3 spacetime.

The equations governing the geodesics in spacetime with the line element $ds^2=g_{ij}dx^idx^j$ can be derived from the lagrangian
\begin{equation}
2 \mathscr{L}=g_{i j} \frac{\mathrm{d} x^{i}}{\mathrm{d} \tau} \frac{\mathrm{d} x^{j}}{\mathrm{d} \tau}\;
\end{equation}
where $\mathrm{\tau}$ is some affine parameter along the geodesic. For time-like geodesics, $\mathrm{\tau}$ may be identified with the proper time of the particle.
Regarding  the metric in the form $ds^2=-f(r)dt^2+1/f(r)dr^2+r^2d\Omega^2$, the lagrangian is
\begin{equation}
\begin{aligned}
\mathcal{L} &=\frac{1}{2}g_{i j} \frac{\mathrm{d} x^{i}}{\mathrm{d} \tau} \frac{\mathrm{d} x^{j}}{\mathrm{d} \tau} \\
&=\frac{1}{2}\left(g_{00} \dot{t}^{2}+g_{11} \dot{r}^{2}+g_{22} \dot{\theta}^{2}+g_{33} \dot{\varphi}^{2}\right) \\
&=\frac{1}{2}\left(-f \dot{t}^{2}+\frac{1}{f} \dot{r}^{2}+r^{2} \dot{\theta}^{2}+r^{2} \sin ^{2} \theta \dot{\varphi}^{2}\right)\;,
\end{aligned}
\end{equation}
where the dot denotes differentiation with respect to $\tau$. The canonical momentum are
\begin{equation}
\begin{array}{ll}
p_{t}=\frac{\partial \mathcal{L}}{\partial \dot{t}}=-f \dot{t}\;, & p_{r}=\frac{\partial \mathcal{L}}{\partial \dot{r}}=\frac{\dot{r}}{f}\;, \\
p_{\varphi}=\frac{\partial \mathcal{L}}{\partial \phi}=r^{2} \sin \theta \dot{\varphi}\;,& p_{\theta}=\frac{\partial \mathcal{L}}{\partial \theta}=r^{2} \dot{\theta}\;,
\end{array}
\end{equation}
and the resulting Hamiltonian is
\begin{equation}
\mathscr{H}=p_{t} \dot{t}+p_{r} \dot{r}+p_{\theta} \theta+p_{\varphi} \dot{\varphi}-\mathscr{L}=\mathscr{L}\;.
\end{equation}

Since Lagrangian doesn't explicitly depend on $\tau$, then the Lagrangian and Hamiltonian are both constant. By rescaling the affine parameter $\tau$, we can arrange that $2\mathscr{L}$ has the value $-1$ for time-like geodesics and $0$ for null geodesics.
Further integral of the motion follows from the equations
\begin{equation}
\frac{d p_{t}}{d \tau}=\frac{\partial \mathscr{L}}{\partial t}=0 \quad \text { and } \quad \frac{d p_{\varphi}}{d \tau}=-\frac{\partial \mathscr{L}}{\partial \varphi}=0\;.
\end{equation}
Then we obtain
\begin{equation}
\label{eqpt}
p_{t}=-f \dot{t}=-E\;,
\end{equation}
and
\begin{equation}
\label{pfai}
p_{\varphi}=r^{2} \sin \theta \dot{\varphi}=constant\;.
\end{equation}

Moreover, from the equation of motion
\begin{equation}
\frac{\mathrm{d} p_{\theta}}{\mathrm{d} \tau}=\frac{\mathrm{d}}{\mathrm{d} \tau}\left(r^{2} \dot{\theta}\right)=-\frac{\partial \mathscr{L}}{\partial \theta}=\left(r^{2} \sin \theta \cos \theta\right)\left(\frac{\mathrm{d} \varphi}{\mathrm{d} \tau}\right)^{2}\;,
\end{equation}
it follows that if we choose to assign the value of $\pi/2$ to $\theta$ when $\dot{\theta}$ is zero, then $\ddot{\theta}$ will also be zero. So $\theta$ will remain constant at the assigned value. We conclude that the geodesic is described in an invariant plane  which we may distinguish by $\theta=\pi/2$. Equation (\ref{pfai}) then gives
\begin{equation}
\label{eqpphi}
p_{\varphi}=r^{2} \frac{\mathrm{d} \varphi}{\mathrm{d} \tau}=\mathrm{constant}=L \quad(\mathrm{say})\;,
\end{equation}
where $L$ denotes the angular momentum about an axis normal to the invariant plane
with $\dot{t}$ and  $\dot{\varphi}$ given by equations  (\ref{eqpt}) and (\ref{eqpphi}). The constancy of the lagrangian gives
\begin{equation}
\label{eqlag}
\frac{E^{2}}{f}-\frac{\dot{r}^{2}}{f}-\frac{L^{2}}{r^{2}}=+1\quad or \quad  0\;.
\end{equation}
\subsection{Null geodesics}

In this subsection we shall restrict ourselves to null geodesics
by considering $r$ as a function of $\varphi$ (instead of $\tau$). From equation (\ref{eqpphi}) and equation (\ref{eqlag}), we obtain
\begin{equation}
\left(\frac{\mathrm{d} r}{\mathrm{d} \varphi}\right)^{2}=\left(E^{2}-\frac{L^2}{r^2}f\right) \frac{r^{4}}{L^{2}}\;.
\end{equation}
Letting
\begin{equation}
u=r^{-1}\;,
\end{equation}
we obtain the basic equation of the problem:
\begin{equation}
\begin{aligned}
\left(\frac{d u}{d \varphi}\right)^{2} &=\frac{E^{2}}{L^{2}}-u^{2} f\;.
\end{aligned}
\end{equation}
Letting $V(u)=-u^2f$, then we obtain $\left(\frac{d u}{d \varphi}\right)^{2}=\frac{E^{2}}{L^{2}}+V(u)$.

For the NBH-3, we have
\begin{equation}
f=1-\frac{r^{2} \sin \left(\frac{2 \beta M}{r^{3}}+\frac{\lambda \beta}{3}\right)}{\beta}\;,
\end{equation}
and
\begin{equation}
V(u)=-u^2\left[1-\frac{ \sin \left(2 \beta Mu^{3}+\frac{\lambda \beta}{3}\right)}{\beta u^2}\right]\;.
\end{equation}
The motion of particles is allowed in the region where $V(u)\geq-\frac{E^2}{L^2}$. Fig.~(8) shows $V(u)$ has many damping oscillations with the increasing of $u$. In particular, when $u\rightarrow \infty$ (or $r\rightarrow 0$), we have $V\rightarrow -\infty$. We conclude that the particle with non-vanishing angular momentum cannot reach the black hole center.

Fig.~(9a) tells us if we release a particle between the cosmological horizon ($u_{\alpha}$) and the black hole event horizon ($u_{\beta}$), it will either bounce off somewhere or  asymptotically spiral around a circle where $-\frac{E^2}{L^2}$ meets exactly the local minimum of $V(u)$. The circle is nothing but the well-known smallest and unstable orbit for massless particles. The radius of the circle is roughly $r=3M$. This is the motion of test particles in the exterior of the black hole.

Now let's consider the motion of test particles in the interior of the black hole. Fig.~(\ref{V2}) tells us when $-\frac{E^2}{L^2}$ meets the local maximum, the massless particle released at the corresponding extreme point will travel in a stable circular orbit. If the particle with the same energy is released at other places, it will be either in a bound orbit or in a so-called semi open-closed interval which we will investigate later.
	
When $-\frac{E^2}{L^2}$ meets the local minimum, the particle released at the corresponding extreme point will travel in an unstable circular orbit and any perturbation will cause it to fly away either inward or outward. Taking $-\frac{E_1^2}{L_1^2}$ as an example, the particle traveling inward will bounce off at ${1}/{u_2}$ and asymptotically spiral around the circle of radius  $r={1}/{u_1}$. On the other hand, the particle traveling outward can escape to infinity. This means the particle can escape from the inside of black hole. We emphasize that the escaping from the black hole is the experience of co-moving observer sitting on the particle. For the observer in infinity, he can not see a particle escaping from black hole due to the presence of black hole event horizon.

In general, for a given value of $\frac{E_0^2}{L_0^2}$, there are a finite number of discrete closed intervals satisfying the condition $V(u)\geq-\frac{E_0^2}{L_0^2}$. Each interval is the subset of $\left[u_{2n+1},u_{2n+2}\right]$ for some $n \in \mathbb{N}$.
The sequence $\left\{ u_n \right\}$ is constructed in this way: $u_{2n+1}$ and $u_{2n+2}$ are separated on both sides of the local minimum of $V(u)$ with $V(u_{2n+1})=V(u_{2n+2}) $. For example, we have $V(u_{1})=V(u_{2}) $ for the interval $\left[u_{1},u_{2}\right]$ and $V(u_{3})=V(u_{4}) $ for the interval $\left[u_{3},u_{4}\right]$ as shown in Fig.~(\ref{V2}). Since the motion is constrained by $V(u)\geq-\frac{E_0^2}{L_0^2}$, we conclude that there are bound orbits which are  oscillating between two spheres with radii $r=1/u_{2n+1}$ and $r=1/u_{2n+2}$, respectively.
As an example, in Fig.~(10a), we plot the wavy orbits in the red interval and green interval (see Fig.~(10b)), respectively. We find the orbits are wavelike squiggle which is significantly different from the precession orbit of Mercury and the Keplerian orbit of our earth.  It is also a bit like the quantum motion of particles.

{There's still a semi open-closed interval in the form of $\left(0,u_0\right]$. $u_0$ satisfying $\frac{E_0^{2}}{L_0^{2}}+V(u_0)=0$. When $\frac{E_1^{2}}{L_1^{2}}<\frac{E_0^{2}}{L_0^{2}}<\frac{E_2^{2}}{L_2^{2}}$, the particle will rebound at $u_0$ and travel to infinity.  Therefore, we conclude that there don't exist stable orbits in the interval $\left(u_2,u_3\right)$, $\left(u_4,u_5\right)$ and so on. By the way, the closed bound orbits are all discrete}.

{In all, the bound orbits can be divided into three categories:}

{1. Stable circular orbits when $-\frac{E^{2}}{L^{2}} $ meets local maximum of $V(u)$ which is equivalent to the conditions:
\begin{equation}
\left\{\begin{array}{l}
V(u)+\frac{E^{2}}{L^{2}}=0\;, \\
V^{\prime}(u)=0\;, \\
V^{\prime \prime}(u)<0\;.
\end{array}\right.
\end{equation}}

{2. Orbits that are asymptotically spiralling around a circle of radius when $-\frac{E^{2}}{L^{2}} $meets local minimum of $V(u)$ which is equivalent to the conditions:
\begin{equation}
\left\{\begin{array}{l}
V(u)+\frac{E^{2}}{L^{2}}=0\;, \\
V^{\prime}(u)=0\;, \\
V^{\prime \prime}(u)>0\;.
\end{array}\right.
\end{equation}}

{3. Wavy orbits that are constrained within two spheres of different radii which are determined by $\frac{E^{2}}{L^{2}}+V(u)=0$. For a given angular momentum $L$, the energy $E$ of the particles would increase with  the increasing of $r=1/u$ which is a bit like the energy levels in atoms.}

\begin{figure}[htbp]
\label{timesmallbeta}
\centering
\label{timesmallbeta}
	\includegraphics[width=8cm,height=6cm]{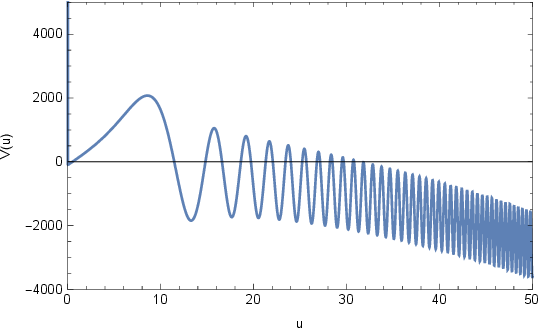}
	\caption{The evolution of $V(u)$ with respect to $u$ when $M=1$, $\lambda=0.0001$, $\beta=0.001$.}
\end{figure}
\begin{figure}[htbp]
	\centering
	\subfigure[ ]{  \label{V1}
		\includegraphics[width=8cm,height=6cm]{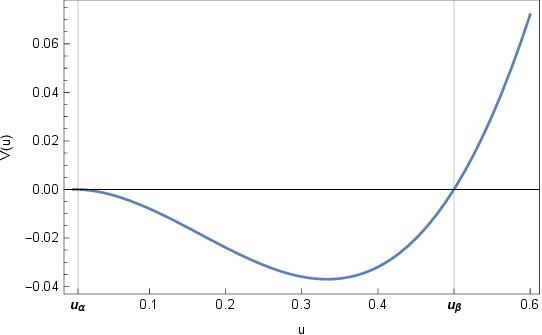}}
	\subfigure[]{\label{V2}
		\includegraphics[width=8cm,height=6cm]{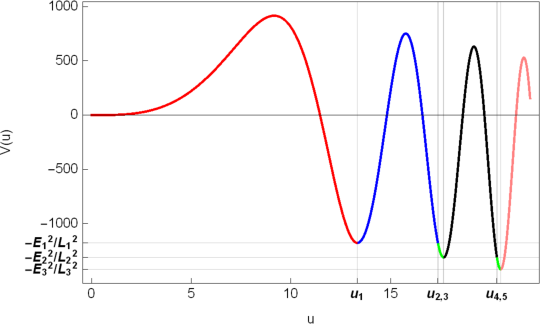}}\\
	\caption{(a) The enlarged portion of Fig.~(8) in the region $0<u<0.6$. $u_\alpha$ represents the cosmological horizon and $u_\beta$ the outermost black hole event horizon. (b) The enlarged portion of Fig.~(8) in the region $0<u<22$.}
	\label{Vuu}
\end{figure}
\begin{figure}[htbp]
	\label{orbit}
	\centering
	\subfigure[]{\label{insidecircle.eps}
		\includegraphics[width=8cm,height=6cm]{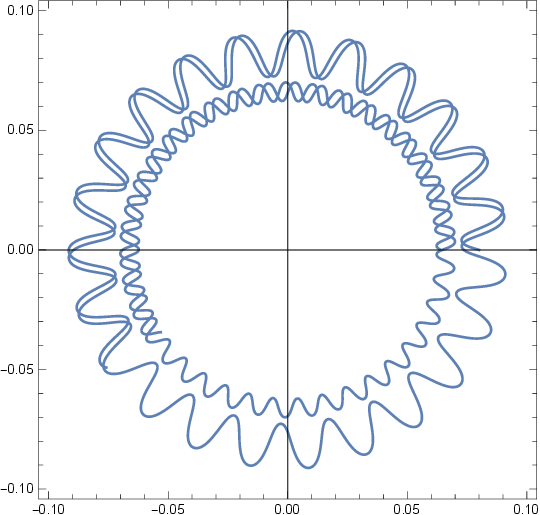}}
	\subfigure[ ]{  \label{KepVu}
		\includegraphics[width=8cm,height=6cm]{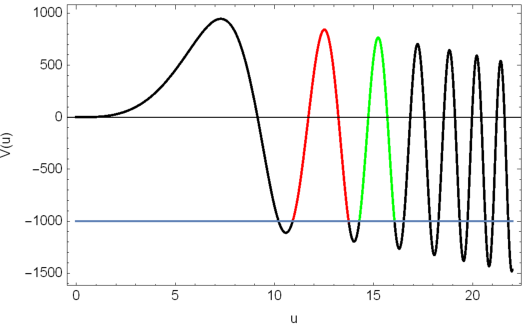}}\\
	\caption{(a) Two wavy orbits inside the black hole when $M=2$, $\lambda=0$, $\beta=0.001$ and $\frac{E^2}{L^2}=1000$. We note that the orbits are not closed. (b) The outer and inner wavy orbits are constrained in the red and green intervals, respectively.}
\end{figure}

\subsection {Time-like geodesics}

Now let's turn to time-like geodesics. Following the procedures of null geodesics, we obtain the equation of motion
\begin{equation}
\left(\frac{d u}{d \varphi}\right)^{2}=\frac{E^{2}}{L^{2}}+V(u)\;,
\end{equation}
where $V(u)=-\left(\frac{1}{L^2}+u^2\right)f$. We see that the potential of $V(u)$ for massive particles, compared to that of massless particles, is given a new term of $-\frac{f}{L^2}$. In other words, we have to take into account the parameters of the particle itself besides the parameters of the black hole. We find the corresponding conclusions on orbits are nearly the same as the null geodesic. Therefore, in the next we shall investigate briefly on the radial motion of particles.
The equations governing the null and time-like geodesics are
 \begin{equation}
 \left(\frac{d r}{d \tau}\right)^{2}=E^2\;, \quad \text { and } \quad \frac{d t}{d \tau}=\frac{E}{f}\;,
 \end{equation}
and
\begin{equation}
\left(\frac{d r}{d \tau}\right)^{2}=E^2-f\;, \quad \text { and } \quad \frac{d t}{d \tau}=\frac{E}{f}\;,
\end{equation}
respectively. For massless particles, we have
\begin{equation}
r=\pm E \tau+\mathrm{constant}_{\pm}\;.
\end{equation}
It indicates that massless particles can reach the center within finite affine ``time'' $\tau$ provided that their angular momentum $L=0$. Of course, for the observer in infinity, it would take infinite coordinate time $t=\int \frac{1}{f}dr$ for the particle to reach the event horizon, needless to say the center of the black hole. We emphasize that the massless particles with non-vanishing angular momentum could never reach the center which has been shown in subsection $\textbf{A}$.
\begin{figure}[htbp]
	\centering
	\includegraphics[width=8cm,height=6cm]{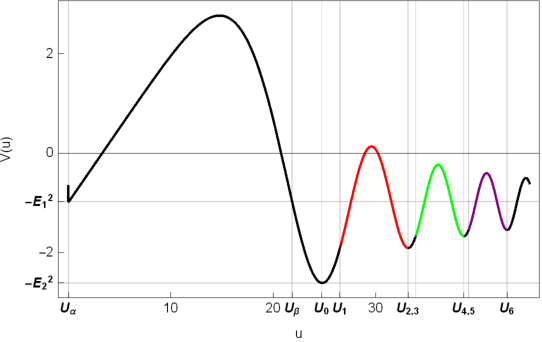}
	\caption{The evolution of $V(u)$ with respect to $u$ when $M=0.15$, $\lambda=0.0001$, $\beta=0.001$.}
	\label{radial}
\end{figure}

{Following the procedure of general cases, we set $V(u)=-f$. Then for a given value of $-{E_0^2}\neq -1$, there are a finite number of discrete closed intervals satisfying the condition $V(u)\geq-{E_0^2}$. Each interval is the subset of $\left[U_{2n+1},U_{2n+2}\right]$ for some $n \in \mathbb{N}$ and the sequence is constructed in the same way as before. We have the conclusion again that there are bound orbits which are oscillating between two points with $r=1/U_{2n+1}$ and $r=1/U_{2n+2}$, respectively.  When $-E^2=-1$, there will be an infinite number of such intervals satisfying the condition $V(u)\geq-\frac{E_0^2}{L_0^2}$. The energy of the bound particle is asymptotically approaching the Planck energy when $u\rightarrow \infty$.

Since the interval has a limit, the energy of the bound orbits should have a upper and lower bounds, which is easy to identify from Fig.~\ref{radial}.
Different from the analysis of non-vanishing angular momentum case, if a particle {is released} between $\left[U_{2n},U_{2n+1}\right]$,
for example, {$U_{\gamma}$} ,between $\left[U_{2},U_{3}\right]$, it will shift inwards and bounce off at the origin, reach {$U_{\gamma}$} and bounce off again, repeat; in other words, we can say that it's a bound particle and constrained in the sphere of radius equal to { $\frac{1}{U_{\gamma}}$}.
	
When $-E^{2}$ meets one local maximum of $-f$, the corresponding position $u$ is stable for the particle. In other words, the particles with energy $E$ can remain there forever. The energy $E$ of the particle is determined by
	\begin{equation}
	\left\{\begin{array}{l}
	V=0\;, \\
	V^{\prime}(r)=0\;, \\
	V^{\prime \prime}(r)<0\;.
	\end{array}\right.
	\end{equation}
They are equivalent to
\begin{equation}
\left\{\begin{array}{l}
E_n^2-\left(1-\frac{r_n^{2} \sin \left(\frac{2 \beta M}{r_n^{3}}+\frac{\lambda \beta}{3}\right)}{\beta}\right)=0\;, \\
3M\beta\cos\left(\frac{2 \beta M}{r_n^{3}}+\frac{\lambda \beta}{3}\right)-r_n^3\sin\left(\frac{2 \beta M}{r_n^{3}}+\frac{\lambda \beta}{3}\right)=0\;, \\
2\left(r_n^6-18\beta^2M^2\right)\sin\left(\frac{2 \beta M}{r_n^{3}}+\frac{\lambda \beta}{3}\right)<0\;.
\end{array}\right.
\end{equation}

It is apparent there are infinite stable energy levels $\{E_n\}$ in the vicinity of black hole center. The energies vary from Planck energy to zero with the increasing of radius. The particles with $E=m_p$ are constrained in the vicinity of center. Then with the increasing of radius, the energy $E_n$ is asymptotically vanishing. We emphasize that all the energy levels are confined in the black hole.

If $-E^2$ meets the local minimum of $V(u)$, for example, $U_{2}$, the particle {released at the corresponding extreme point} will stay there and the point is unstable to any perturbations. The particles perturbed shifting inwards will reach the origin and bounce off, then asymptotically approach the radius equal to ${1}/{U_{2}}$. The particles perturbed shifting outwards will bounce off at ${1}/{U_{1}}$ and asymptotically approach the radius equal to ${1}/{U_{2}}$. 	

 Finally, let's consider the motion of test particle which is released between the cosmological horizon and the black hole event horizon. When $E^2<E_{1}^2$, it will be constrained in $u<U_{\alpha}$ or $U_{\alpha}<u<U_{\beta}$. When $E_{1}^2<E^2<E_{2}^2$, the particle will never enter the sphere of radius equal to ${{1}/{U_{0}}}$. It will bounce off where $E^2-f$ vanishes, and then fly to infinity. If the energy $E$ of the particle satisfies  $E^2\gg E_2^2$, it will be able to arrive the center and then rebound to infinity. Now we see that, different from massless particles, massive particles cannot always arrive at the center.}

\section{4-dimensional nonsingular universes}

In this section, we shall show there are a large number of nonsingular universe solutions in the regularized Lovelock gravity. To this end, we start from the n-dimensional Friedmann equation in Lovelock gravity which is derived by Deruelle and Farina-Busto \cite{far:1990}
\begin{eqnarray}\label{frw}
\sum_{p=0}^{N}\frac{\alpha_p}{2}H^{2p}\frac{\left(n-1\right)!}{\left(n-2p-1\right)!}=\kappa \rho\;,
\end{eqnarray}
where $H$ is the Hubble parameter and $\rho$ is the energy density of universe. Here a spatially flat universe is assumed. The factorial function in the denominator tells us we must have
\begin{eqnarray}
p\leq\frac{n-1}{2}\;.
\end{eqnarray}
When $n=4$, we are left with only $p=0$ and $p=1$. Eq.~(\ref{frw}) is then reduced to
\begin{eqnarray}
\frac{\alpha_0}{2}+3H^{2}=\kappa \rho\;,
\end{eqnarray}
provided that
\begin{eqnarray}
\alpha_1=1\;.
\end{eqnarray}
It is exactly the Friedmann equation in General Relativity. Now $\alpha_0$ plays the role of cosmological constant.
We see the higher orders $p\geq 2$ of Lovelock tensors make no contributions to the equations of motion.
However, it is not the case in regularized Lovelock gravity. To show this point, we make the regularization as follows \cite{CasalinoCRV}
\begin{eqnarray}
\alpha_p\rightarrow 2{c}_p\frac{\left(n-2p-1\right)!}{\left(n-1\right)!}\;.
\end{eqnarray}
Namely, the coupling constants $\alpha_p$ are replace with $c_p$. Then the regularized Friedmann equation turns out to be
\begin{eqnarray}\label{frwr}
\sum_{p=0}^{N}c_p H^{2p}=\frac{8\pi}{3}\rho\;,
\end{eqnarray}
where $c_p$ are understood as the regularized coupling constants and $\kappa=8\pi/3$ is adopted. Now all the orders of Lovelock tensor contribute to the equations of motion regardless of the dimension of spacetime. Since the order $N$ of lovelock tensors can be arbitrarily large, we can let $N\rightarrow \infty$. In order that General relativity is covered in the theory, we require that
\begin{eqnarray}\label{frwr}
c_0=-\frac{\lambda}{3}\;,\ \ \ \ c_1=1\;,
\end{eqnarray}
$\lambda$ is the cosmological constant.

Same as in section III, we introduce function $P(\psi)$ as follows
\begin{eqnarray}
P\left(\psi\right)\equiv\sum_{p=0}^{\infty}c_p \psi^{p}\;,
\end{eqnarray}
where $\psi$ is defined by
\begin{eqnarray}
\psi\equiv H^2\;.
\end{eqnarray}
We obtain the regularized Friedmann equation in Lovelock frame
\begin{eqnarray}\label{rfr}
P\left(\psi\right)=\frac{8\pi}{3}\rho\;.
\end{eqnarray}
Comparing it with Eq.~(\ref{MEq}), we find
\begin{eqnarray}\label{rfr}
M=\frac{4\pi}{3}\rho r^3\;,
\end{eqnarray}
a very delicate result.

Solving for $\psi$ or $H^2$, we obtain the regularized Friedmann equation in Einstein frame
\begin{eqnarray}
H^2=P^{-1}\left(\rho\right)\;.
\end{eqnarray}
The energy density of the universe is often parameterized as
\begin{eqnarray}
\rho=\frac{\rho_{s_0}}{a^6}+\frac{\rho_{r_0}}{a^4}+\frac{\rho_{d_0}}{a^3}\;,
\end{eqnarray}
which denote stiff matter, radiation matter and dark matter, respectively. For the present universe, we have $\lambda\simeq\rho_{d_0}\gg\rho_{r_0}\gg\rho_{s_0}$.  In the next subsections, we shall present several nonsingular universe solutions.
\subsection{nonsingular universe-1}
We select the same function
\begin{equation}
P\left(\psi\right)=-\frac{\lambda}{3}+\frac{1}{\beta}\left[\left(1+\frac{\beta\psi}{\gamma}\right)^{\gamma}-1\right]\,,
\end{equation}
as in the first nonsingular black hole solution and substitute it into the regularized Friedmann equation Eq.~(\ref{rfr}). Then the Friedmann equation in Einstein frame is derived
\begin{equation}\label{Lmetric}
H^2=\frac{\gamma}{\beta}\left[\left(1+\frac{8\pi\beta\rho}{3}+\frac{\lambda\beta}{3}\right)^{\frac{1}{\gamma}}-1\right]\;.
\end{equation}
We assume $\lambda>0,\ \beta>0$ and $\gamma<0$. When $\beta\rightarrow 0$, it restores to the Friedmann equation in General Relativity.
When $\rho\rightarrow \infty$, we have the de Sitter solution
\begin{equation}\label{Lmetric}
a\propto e^{ct\sqrt{\frac{-\gamma}{\beta}}}\;,
\end{equation}
where $c$ is the speed of light. It is generally conjectured that the cosmic inflation starts from the Planck length, $l_p\sim 10^{-35}m$. Therefore, $\beta$ is the order of $\beta\sim l_p^2$.
On the other hand, when $\rho\rightarrow 0$, we also have the de Sitter solution
\begin{equation}\label{DE}
a\propto e^{ct\sqrt{1-\left(1+\frac{\lambda\beta}{3}\right)^{\frac{1}{\gamma}}}\sqrt{\frac{-\gamma}{\beta}}}\;.
\end{equation}
We know the present-day cosmological constant is in the order of inverse of square of the present-day Hubble length $\lambda\sim L_0^{-2}\sim \left(\frac{H_0}{c}\right)^2\simeq 1.2\times 10^{26} m $. So we have
\begin{equation}
\frac{\lambda\beta}{3}\simeq\frac{l_p^2 L_0^{-2}}{3}\simeq 10^{-120}\ll 1\;.
\end{equation}
Therefore, Eq.~(\ref{DE}) can be approximately written as
\begin{equation}
a\propto e^{\sqrt{\frac{\lambda}{3}}t}\;.
\end{equation}
In all, this is a nonsingular universe solution beginning in a de Sitter phase and ending in another de Sitter phase. The energy densities of the two phases are $\rho_1\sim \lambda$ and $\rho_2\sim 1/\beta$, respectively. Then the difference of vacuum energy densities
for the two phase is about $\rho_1/\rho_2\sim 10^{-120}$, namely, $120$ orders of magnitude.

\subsection{nonsingular universe-2}
We select the same function
\begin{equation}\label{Ppsi}
P\left(\psi\right)=-\frac{\lambda}{3}-\frac{1}{\beta}\ln\left(1-{\beta}{\psi}\right)\,,
\end{equation}
as in the second nonsingular black hole solution and substitute it into the regularized Friedmann equation Eq.~(\ref{rfr}). Then the Friedmann equation in Einstein frame is derived
\begin{equation}\label{Lmetric}
H^2=\frac{1}{\beta}\left(1-e^{-\frac{8\pi\beta\rho}{3}-\frac{\lambda\beta}{3}}\right)\;.
\end{equation}
Same as the first solution, we assume $\beta\sim l_p^2$ and $\lambda\sim L_0^{-2}$. When $\beta\rightarrow 0$, it restores to the Friedmann equation in General Relativity.
When $\rho\rightarrow \infty$, we have the de Sitter solution
\begin{equation}\label{Lmetric}
a\propto e^{\sqrt{\frac{1}{\beta}}t}\;.
\end{equation}
On the other hand, when $\rho\rightarrow 0$, it also gives a de Sitter solution
\begin{equation}\label{Lmetric}
a\propto e^{\sqrt{\frac{1}{\beta}\left(1-e^{-\frac{\lambda\beta}{3}}\right)}t}\simeq e^{\sqrt{\frac{\lambda}{3}}t}\;.
\end{equation}
Therefore, this remains a nonsingular universe solution beginning in a de Sitter phase and ending in another de Sitter phase. The difference of vacuum energy densities
for the two phases is about $120$ orders of magnitude.

\subsection{nonsingular universe-3}
\begin{figure}[htbp]
	\centering
	\includegraphics[width=8cm,height=6cm]{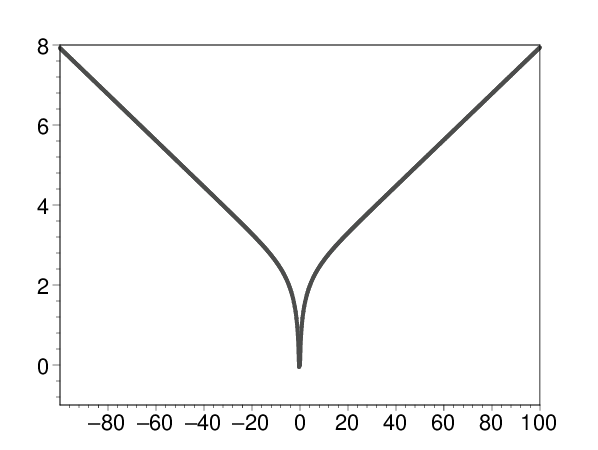}
	\caption{The evolution of $\ln{a}$ with respect to cosmic time $t$. The universe starts from a bounce and ends in a de Sitter expansion.}\label{bounce}
\end{figure}

We select the same function
\begin{equation}\label{Ppsi}
P\left(\psi\right)=-\frac{\lambda}{3}+\frac{1}{\beta}\arcsin\left(\beta\psi\right)\,,
\end{equation}
as in the third nonsingular black hole solution and substitute it into the regularized Friedmann equation Eq.~(\ref{rfr}). Then the Friedmann equation in Einstein frame is derived
\begin{equation}\label{Lmetric}
H^2=\frac{\sin\left(\frac{8\pi\beta\rho}{3}+\frac{\lambda\beta}{3}\right)}{\beta}\;.
\end{equation}
When $\beta\rightarrow 0$, it restores to the standard Friedmann equation in General Relativity. When $\rho\rightarrow 0$ and taking into  account of $\lambda\beta\ll 1$, we obtain the de Sitter universe
\begin{equation}\label{Lmetric}
a\propto e^{\sqrt{\frac{\sin\left(\frac{\lambda\beta}{3}\right)}{\beta}}t}\simeq e^{\sqrt{\frac{\lambda}{3}}t}\;.
\end{equation}
The vacuum energy density in this phase is the order of $\lambda\sim H_0^2$. The matter density $\rho$ can not be arbitrarily large. In fact, when $\rho\rightarrow 3/(8\beta)$, we have $\frac{8\pi\beta\rho}{3}+\frac{\lambda\beta}{3}=\pi+\lambda\beta/3\simeq \pi$ such that $\dot{a}=0$ and
\begin{equation}
\dot{H}\propto -\cos{\left(\frac{8\pi\beta\rho}{3}+\frac{\lambda\beta}{3}\right)}\left(\rho+p\right)\simeq \rho+p>0\;.
\end{equation}
Then a bounce universe is achieved. In the phase of bounce, the energy density is the order of $\rho\sim 1/\beta=\rho_p$, namely, the Planck energy density.
In Fig.~\ref{bounce} we plot the evolution of $\ln {a}$ with cosmic time $t$. We put $\beta=0.1,\ \lambda=0.7\cdot 8\pi,\ \rho_{s0}=10^{-6},\  \rho_{r0}=10^{-4},\ \rho_{d0}=0.3$. It shows that the universe starts from a bounce and then undergoes the stiff matter dominated epoch, the radiation matter dominated epoch, the dark matter dominated epoch, ends lastly in a de Sitter expansion.

Finally, one can find other nonsingular universe solutions by using the input-functions in Eq.~(\ref{functions}). By and large, these solutions are divided into two types. The first type states that the universe starts from a de Sitter phase and ends in another de Sitter phase.  The second type states that the universe starts from a bounce and ends in a de Sitter expansion.

\section{conclusion and discussion}
In conclusion, we have shown there are a large number of nonsingular black holes and nonsingular universes in the regularized Lovelock gravity in four dimensional spacetime. To obtain the nonsingular black holes and nonsingular universes in the framework of General Relativity, one must resort to exotic physical sources which generally violate the energy conditions. This is required by the well-known singularity theorems \cite{penrose:1965}. However, it is not the case in our considerations. It does not need to introduce any physical sources in our calculations. In this sense, the solutions are built purely in vacuum.
We present the nonsingular black hole and nonsingular universe with the same function of $P(\psi)$. But the corresponding gravitational theories are different because the coupling constants $c_p$ are different.

It is found that there are in general three horizons for NBH-1 (nonsingular black hole-1) and NBH-2 (nonsingular black hole-2). They are cosmic horizon, event horizon and inner Cauchy horizon, respectively. However, for NBH-3 (nonsingular black hole-3), there can exist a fairly large number of horizons. These black holes are all asymptotically de Sitter or anti-de Sitter in space. In the absence of cosmological constant, they are asymptotically Minkowski. Finally, when $r\rightarrow 0$, the spacetime reduces to Minkowski and the black hole central singularity is erased. In order to understand the internal structure of black holes, we study the geodesic motions of massless and massive particles. We find there are three types of orbits inside the black hole.

1. When $-{E^{2}}/{L^{2}} $ meets one local maximum of $V(u)$, there would exist a stable and circular orbit. Since there are infinite local maximum on the potential $V$, the number of stable and circular orbit is also infinite.

2. When $-{E^{2}}/{L^{2}} $ meets one local minimum of $V(u)$, there would exist asymptotically spiralling circular orbit. The number of asymptotically spiralling circular orbits is also infinite.

3. When $-{E^{2}}/{L^{2}} $ is between one local maximum and one local minimum of $V(u)$, there would exist a number of wavy orbits. Every wavy orbit is confined between two spheres of different radius. For fixed $-E^2/L^2$, the number of wavy orbits is finite.
By considering the radial motion of massive particles, we find there exist infinite energy levels inside the black hole.

Finally, we find that both NU-1 (nonsingular universe-1) and  NU-2 (nonsingular universe-2) are asymptotically de Sitter whether for $\rho\rightarrow 0$ or for $\rho\rightarrow \infty$. This means they are created in a de Sitter phase and are ended in another de Sitter phase.  The difference of vacuum energy densities
for the two phases is about $120$ orders of magnitude. Thus it may be helpful to understand the cosmological constant problem.
For NU-3 (nonsingular universe-3), it is also asymptotically de Sitter in distant future. But it is created from a bounce while not a de Sitter phase.

\section*{Acknowledgments}
This work is partially supported by the Strategic Priority Research Program ``Multi-wavelength Gravitational Wave Universe'' of the
CAS, Grant No. XDB23040100 and the NSFC under grants 11633004, 11773031.

\newcommand\ARNPS[3]{~Ann. Rev. Nucl. Part. Sci.{\bf ~#1}, #2~ (#3)}
\newcommand\AL[3]{~Astron. Lett.{\bf ~#1}, #2~ (#3)}
\newcommand\AP[3]{~Astropart. Phys.{\bf ~#1}, #2~ (#3)}
\newcommand\AJ[3]{~Astron. J.{\bf ~#1}, #2~(#3)}
\newcommand\APJ[3]{~Astrophys. J.{\bf ~#1}, #2~ (#3)}
\newcommand\APJL[3]{~Astrophys. J. Lett. {\bf ~#1}, L#2~(#3)}
\newcommand\APJS[3]{~Astrophys. J. Suppl. Ser.{\bf ~#1}, #2~(#3)}
\newcommand\JHEP[3]{~JHEP.{\bf ~#1}, #2~(#3)}
\newcommand\JMP[3]{~J. Math. Phys. {\bf ~#1}, #2~(#3)}
\newcommand\JCAP[3]{~JCAP {\bf ~#1}, #2~ (#3)}
\newcommand\LRR[3]{~Living Rev. Relativity. {\bf ~#1}, #2~ (#3)}
\newcommand\MNRAS[3]{~Mon. Not. R. Astron. Soc.{\bf ~#1}, #2~(#3)}
\newcommand\MNRASL[3]{~Mon. Not. R. Astron. Soc.{\bf ~#1}, L#2~(#3)}
\newcommand\NPB[3]{~Nucl. Phys. B{\bf ~#1}, #2~(#3)}
\newcommand\CMP[3]{~Comm. Math. Phys.{\bf ~#1}, #2~(#3)}
\newcommand\CQG[3]{~Class. Quantum Grav.{\bf ~#1}, #2~(#3)}
\newcommand\PLB[3]{~Phys. Lett. B{\bf ~#1}, #2~(#3)}
\newcommand\PRL[3]{~Phys. Rev. Lett.{\bf ~#1}, #2~(#3)}
\newcommand\PR[3]{~Phys. Rep.{\bf ~#1}, #2~(#3)}
\newcommand\PRd[3]{~Phys. Rev.{\bf ~#1}, #2~(#3)}
\newcommand\PRD[3]{~Phys. Rev. D{\bf ~#1}, #2~(#3)}
\newcommand\RMP[3]{~Rev. Mod. Phys.{\bf ~#1}, #2~(#3)}
\newcommand\SJNP[3]{~Sov. J. Nucl. Phys.{\bf ~#1}, #2~(#3)}
\newcommand\ZPC[3]{~Z. Phys. C{\bf ~#1}, #2~(#3)}
\newcommand\IJGMP[3]{~Int. J. Geom. Meth. Mod. Phys.{\bf ~#1}, #2~(#3)}
\newcommand\IJMPD[3]{~Int. J. Mod. Phys. D{\bf ~#1}, #2~(#3)}
\newcommand\GRG[3]{~Gen. Rel. Grav.{\bf ~#1}, #2~(#3)}
\newcommand\EPJC[3]{~Eur. Phys. J. C{\bf ~#1}, #2~(#3)}
\newcommand\PRSLA[3]{~Proc. Roy. Soc. Lond. A {\bf ~#1}, #2~(#3)}
\newcommand\AHEP[3]{~Adv. High Energy Phys.{\bf ~#1}, #2~(#3)}
\newcommand\Pramana[3]{~Pramana.{\bf ~#1}, #2~(#3)}
\newcommand\PTP[3]{~Prog. Theor. Phys{\bf ~#1}, #2~(#3)}
\newcommand\APPS[3]{~Acta Phys. Polon. Supp.{\bf ~#1}, #2~(#3)}
\newcommand\ANP[3]{~Annals Phys.{\bf ~#1}, #2~(#3)}


\begin{thebibliography}{99}

\bibitem{ll:1971}D. Lovelock, J. Math. Phys. {\bf 12}(1971)498.
\bibitem{ll:1972}D. Lovelock, J. Math. Phys. {\bf 13}(1972)874.


\bibitem{nonminimal} E. Abdalla, R. A. Konoplya, C. Molina, Phys. Rev. D
{\bf 72} (2005) 084006; Daniela D. Doneva, Kalin V. Staykov,
Stoytcho S. Yazadjiev, Phys. Rev. D {\bf 99} (2019) 104045;
Roman A. Konoplya, Thomas Pappas, Alexander Zhi-
denko, Phys. Rev. D {\bf 101} (2020) 044054; M. Grses, Gen.
Rel. Grav. {\bf 40} (2008) 1825.

\bibitem{non-l}Y. Zhong, D. Saez-Chillon Gomez, Symmetry, {\bf 10}
(2018) 170; R. Rashidi, F. Ahmadi, M.R. Setare, As-
trophys Space Sci {\bf 363} (2018) 196; A. De Felice(Louvain
U., CP3), S. Tsujikawa, Phys. Lett. B {\bf 675} (2009) 1.

\bibitem{glavan:2020} D. Glavan and C. Lin, Einstein-Gauss-Bonnet gravity in 4-dimensional space-time, Phys. Rev. Lett. 124 (2020) 081301
[1905.03601].

\bibitem{CasalinoCRV}
A. Casalino, A. Colleaux, M. Rinaldi, and S. Vicentini,
arXiv:2003.07068 [gr-qc].


\bibitem{tom:2011} Y. Tomozawa, Quantum corrections to gravity, 1107.1424.
\bibitem{cog:2013} G. Cognola, R. Myrzakulov, L. Sebastiani and S. Zerbini, Einstein gravity with Gauss-Bonnet entropic corrections,
Phys. Rev. D 88 (2013) 024006 [1304.1878].









\bibitem{KumarGhoshMaharaj}
R. Kumar and S. G. Ghosh,
arXiv:2003.08927 [gr-qc]; S. G. Ghosh and S.D. Maharaj,
arXiv:2003.09841 [gr-qc]; S. G. Ghosh and R. Kumar,
arXiv:2003.12291 [gr-qc]; A. Kumar and R. Kumar,
arXiv:2003.13104 [gr-qc]; A. Kumar and S. G. Ghosh,
arXiv:2004.01131 [gr-qc]; S. G. Ghosh and S. D. Maharaj,
arXiv:2004.13519 [gr-qc].

\bibitem{SinghGM}
D. V. Singh, S. G. Ghosh, and S. D. Maharaj,
arXiv:2003.14136 [gr-qc].

\bibitem{DonevaYazadjiev}
D. D. Doneva and S. S. Yazadjiev,
arXiv:2003.10284 [gr-qc].

\bibitem{JusufiBG}
K. Jusufi, A. Banerjee, and S. G. Ghosh,
arXiv:2004.10750 [gr-qc].

\bibitem{GeSin}
X. H. Ge and S. J. Sin,
arXiv:2004.12191 [hep-th].

\bibitem{Liu14267}
P. Liu, C. Niu, X. B. Wang, and C. Y. Zhang,
arXiv:2004.14267 [gr-qc].

\bibitem{Yang14468}
K. Yang, B. M. Gu, S. W. Wei, and Y. X. Liu,
arXiv:2004.14468 [gr-qc].

\bibitem{MaLu}
L. Ma and H. Lu,
arXiv:2004.14738 [gr-qc].


\bibitem{KonoplyaZinhailoZhidenko}
R. A. Konoplya and A. F. Zinhailo,
arXiv:2003.01188 [gr-qc]; R.A. Konoplya and A. Zhidenko,
arXiv:2003.12492 [gr-qc].

\bibitem{Churilova}
M. S. Churilova,
arXiv:2004.00513 [gr-qc];
arXiv:2004.14172 [gr-qc].

\bibitem{Mishra}
A. K. Mishra,
arXiv:2004.01243 [gr-qc].

\bibitem{LiWY}
S. L. Li, P. X. Wu, and H. W. Yu,
arXiv:2004.02080 [gr-qc].

\bibitem{ZhangZLG}
C. Y. Zhang, S. J. Zhang, P. C. Li, and M. Y. Guo,
arXiv:2004.03141 [gr-qc].

\bibitem{AragonBGV}
A. Arag\'{o}n, R. B\'{e}car, P.A. Gonz\'{a}lez, and Y. V\'{a}squez
arXiv:2004.05632 [gr-qc].

\bibitem{MalafarinaTD}
D. Malafarina, B. Toshmatov, and N. Dadhich,
arXiv:2004.07089 [gr-qc].

\bibitem{Cuyubamba09025}
M. A. Cuyubamba,
arXiv:2004.09025 [gr-qc].

\bibitem{LiuNZ}
P. Liu, C. Niu, and C.Y. Zhang,
arXiv:2004.10620 [gr-qc].

\bibitem{Devi14935}
S. Devi, R. Roy, and S. Chakrabarti,
arXiv:2004.14935 [gr-qc].

\bibitem{GuoLi}
M. Y. Guo and P. C. Li,
arXiv:2003.02523 [gr-qc].

\bibitem{WeiLiu07769}
S. W. Wei and Y. X. Liu,
arXiv:2003.07769 [gr-qc].

\bibitem{ZhangWeiLiu}
Y. P. Zhang, S. W. Wei, and Y.X. Liu,
arXiv:2003.10960 [gr-qc].

\bibitem{Heydari-Fard}
Mohaddese Heydari-Fard, Malihe Heydari-Fard, and H. R. Sepangi,
arXiv:2004.02140 [gr-qc].

\bibitem{RayimbaevATA}
J. Rayimbaev, A. Abdujabbarov, B. Turimov, and F. Atamurotov,
arXiv:2004.10031 [gr-qc].

\bibitem{ZengZZ}
X. X. Zeng, H. Q. Zhang, and H. B. Zhang,
arXiv:2004.12074 [gr-qc].


\bibitem{LiuZW}
C. Liu, T. Zhu, and Q. Wu,
arXiv:2004.01662 [gr-qc].

\bibitem{KumaraRHAA}
A. N. Kumara, C.L. Ahmed Rizwan, K. Hegde, M. S. Ali, and K. M. Ajith,
arXiv:2004.04521 [gr-qc].

\bibitem{IslamKG}
S. U. Islam, R. Kumar, and S. G. Ghosh,
arXiv:2004.01038 [gr-qc].

\bibitem{JinGL}
X. H. Jin, Y. X. Gao, and D. J. Liu,
arXiv:2004.02261 [gr-qc].

\bibitem{Kumar12970}
R. Kumar, S.U. Islam, and S. G. Ghosh,
arXiv:2004.12970 [gr-qc].


\bibitem{HegdeKA}
K. Hegde, A.N. Kumara, C. L. Ahmed Rizwan, K. M. Ajith, and M. S. Ali,
arXiv:2003.08778 [gr-qc].

\bibitem{SinghS}
D.V. Singh and S. Siwach,
arXiv:2003.11754 [gr-qc].

\bibitem{ZhangLG}
C. Y. Zhang, P. C. Li, and M. Y. Guo,
arXiv:2003.13068 [hep-th].

\bibitem{Mansoori}
S. A. H. Mansoori,
arXiv:2003.13382 [gr-qc].

\bibitem{WeiL14275}
S. W. Wei and Y. X. Liu,
arXiv:2003.14275 [gr-qc].

\bibitem{Konoplya02248}
R. A. Konoplya and A. F. Zinhailo,
arXiv:2004.02248 [gr-qc].

\bibitem{PanahJafarzade}
B. E. Panah and K. Jafarzade,
arXiv:2004.04058 [gr-qc].

\bibitem{YangWCYW}
S. J. Yang, J. J. Wan, J. Chen, J. Yang, and Y.Q. Wang,
arXiv:2004.07934 [gr-qc].

\bibitem{Ying}
S.X. Ying,
arXiv:2004.09480 [gr-qc].




\bibitem{LuPangMao}
Z. C. Lin, K. Yang, S. W. Wei, Y. Q. Wang, and Y. X. Liu, arXiv:2006.07913 [gr-qc]; H. Lu and Y. Pang,
arXiv:2003.11552 [gr-qc]; H. Lu and P. J. Mao,
arXiv:2004.14400 [hep-th].

\bibitem{Kobayashi}
T. Kobayashi,
arXiv:2003.12771 [gr-qc].

\bibitem{Fernandes08362}
P. G. S. Fernandes, P. Carrilho, T. Clifton, and D. J. Mulryne,
arXiv:2004.08362 [gr-qc].

\bibitem{HennigarKMP}
R. A. Hennigar, D. Kubiznak, R. B. Mann, and C. Pollack,
arXiv:2004.09472 [gr-qc].

\bibitem{BonifacioHJ}
J. Bonifacio, K. Hinterbichler, and L. A. Johnson,
arXiv:2004.10716 [hep-th].
\bibitem{qiao}X. Qiao,  O. Liang, D.  Wang, Q. Pan and J, Jing, arXiv:2005.01007 [hep-th]

\bibitem{Ai2020}
W.Y. Ai,
arXiv:2004.02858 [gr-qc].

\bibitem{GursesST}
M. Gurses, T.C. Sisman, and B. Tekin,
arXiv:2004.03390 [gr-qc].

\bibitem{Mahapatra2020}
S. Mahapatra,
arXiv:2004.09214 [gr-qc].

\bibitem{Shu09339}
F.W. Shu,
arXiv:2004.09339 [gr-qc].

\bibitem{TianZhu}
S.X. Tian and Z.H. Zhu,
arXiv:2004.09954 [gr-qc].

\bibitem{ArrecheaDJ}
J. Arrechea, A. Delhom, and A. Jim\'{e}nez-Cano,
arXiv:2004.12998 [gr-qc].





\bibitem{bh-5} E. Ayon-Beato, A. Garcfa, Phys.Rev.Lett. 80 (1998) 5056; Phys.Lett. B464 (1999) 25;
\bibitem{bh-6}  C. Lechner, S. Husa, P. C. Aichelburg, Phys.Rev. D62 (2000) 044047;
\bibitem{bh-7}  M. Cataldo, A. Garcia, Phys.Rev. D61 (2000) 084003;
\bibitem{bh-8}  K. A. Bronnikov, gr-qc/0006014 ;
\bibitem{bh-9} J. Bardeen, in Proceedings of GR5, Tiflis, U.S.S.R. (1968).
\bibitem{bh-10}  E. Ayon-Beato, Asymptotic Behavior of Scalar Fields Coupled to Gravity, Graduate Dissertation, Faculty of Physics,
Havana Univ. (1993).
\bibitem{bh-11}  Borde, A., Phys. Rev., D50, 3392 (1994).
\bibitem{bh-12}  Barrabes, C., Frolov, V.P., Phys. Rev., D53, 3215 (1996).
\bibitem{bh-13}  Mars, M., Martin-Prats, M.M., Senovilla, J.M.M., Class. Quant. Grav., 13, L51 (1996).
\bibitem{bh-14}  Cabo, A., Ayon¨CBeato, E., Int. J. Mod. Phys., A14, 2013 (1999).
\bibitem{bh-15}  Borde, A., Phys. Rev., D55, 7615 (1997).
\bibitem{bh-16}  Ayon-Beato, E., Garc?a, A., Phys. Rev. Lett., 80, 5056 (1998).
\bibitem{bh-17}  Magli, G., ¡°Physically Valid Black Hole Interior Models,¡± gr-qc/9706083 (1997).
\bibitem{bh-18} P. A. Cano and A. Murcia, arXiv:2006.15149 [hep-th]; I. Dymnikova, Gen. Rel. Grav. 24, 235 (1992); Class. Quant. Grav. 21, 4417 (2004) [gr-qc/0407072]; K. A. Bronnikov, Phys. Rev. D 63, 044005 (2001) [gr-
qc/0006014]; K. A. Bronnikov and J. C. Fabris, Phys. Rev. Lett. 96, 251101 (2006) [gr-qc/0511109]; W. Berej, J. Matyjasek,
D. Tryniecki and M. Woronowicz, Gen. Rel. Grav. 38, 885 (2006) [hep-th/0606185]; A. H. Chamseddine, V. Mukhanov, Eur. Phys. J. C 77, 183(2017); 
I. Dymnikova, M.Khlopov, Int. J. Mod. Phys. D24 (2015).
\bibitem{hayward:2006}S. A. Hayward, Phys. Rev. Lett. 96, 031103 (2006) [gr-qc/0506126].


\bibitem{gao:2018}C. Gao, Y. Lu, S. Yu and Y. G. Shen, \PRD{97}{104013}{2018}; K. A. Bronnikov, I. G. Dymnikova and E. Galaktionov, Class. Qusantum Gravity. {\bf 29},095025(2012); arXiv: 1204.0534;
S. V. Bolokhov, K. A. Bronnikov and M. V. Skvortsova, Class. Qusantum Gravity. {\bf 29},245006(2012); arXiv: 1208.4619;
K. A. Bronnikov, K. A. Baleevskikh and M. V. Skvortsova, arXiv: 1708.02324;
S. Nojiri and S. D. Odintsov, arXiv:1708.05226 [hep-th].



\bibitem{bh-19} J. H. Horne and G. T. Horowitz, Nucl. Phys. B368, 444
(1992).

\bibitem{nicolini:2006} P. Nicolini, A. Smailagic, E. Spallucci, Phys. Lett. B 632, 547(2006).
\bibitem{nicolini:2007} S. Ansoldi, P. Nicolini, A. Smailagic, E. Spallucci, Phys. Lett. B645, 261(2007)



\bibitem{tsey:1995}Tseytlin, A.A., Phys. Lett., B363, 223 (1995); S. Shankaranarayanan and N. Dadhich, Int. J. Mod. Phys. D 13 (2004) 1095-1104.

\bibitem{gue:2020}M. Guerrero, D. Rubiera-Garcia, Phys. Rev. D 102, 024005 (2020)


\bibitem{uni-1}S. Tsujikawa, R. Brandenberger, F. Finelli, Phys. Rev. D. 2002, 66, 083513,1-20.
\bibitem{uni-2}T. Biswas, A. Mazumdar, W. Siegel, JCAP. 2006, 0603,
009.
\bibitem{uni-3}V. F. Mukhanov, R. H. Brandenberger, Phys. Rev. Lett. 1992, 68, 1969-1972
\bibitem{lag-4}Y. F.  Cai, E. N. Saridakis, JCAP. 2009, 0910,
020
\bibitem{bran-5}Y. Shtanov, V. Sahni, Phys. Lett. B. 2003, 557, 1-6.
\bibitem{report-6} M. Novello, S. E. P. Bergliaffa, Phys. Rept. 2008, 463, 127-213.

\bibitem{lan:1932}C. Lanczos, Z. Phys. 73 (1932)147-168.
\bibitem{lan:1938}C. Lanczos, Ann. Math. 39 (1938)842-850.
\bibitem{bri:1997}C. C. Briggs, gr-qc/9703074




\bibitem{Boulware:1985wk}
  D.~G.~Boulware and S.~Deser,
  Phys.\ Rev.\ Lett.\  {\bf 55}, 2656 (1985)


\bibitem{Wheeler}
  J.~T.~Wheeler, Nucl.\ Phys.\ B {\bf 273}, 732 (1986); Nucl.\ Phys.\ B {\bf 268}, 737 (1986)


\bibitem{Wiltshire:1985us}
  D.~L.~Wiltshire, Phys.\ Lett.\  {\bf 169B}, 36 (1986)

 \bibitem{Cai:2001dz}
   R.~G.~Cai,
  Phys.\ Rev.\ D {\bf 65}, 084014 (2002)

\bibitem{far:1990}N. Deruelle and L. Farina-Busto, \PRD{41}{3696}{1990}.

\bibitem{penrose:1965}R. Penrose, Phys. Rev. Lett. 14, 57 (1965); S. W. Hawking,
Proc. R. Soc. London A300, 182 (1967); S. W. Hawking, R. Penrose, Proc. R. Soc. London A314, 529 (1970).




\bibitem{will:1985} B. F. Schutz and C. M. Will, Astrophys. J. Lett. Ed. 291, L33 (1985).
\bibitem{will:1987} S. Iyer and C. M. Will, Phys. Rev. D 35, 3621 (1987).
\bibitem{iyer:1987} S. Iyer, Phys. Rev. D 35, 3632 (1987).



\bibitem{quasi:99} K. D. Kokkotas and B. F. Schutz, Phys. Rev. D 37 (1988) 3378; E. Seidel and S. Iyer, Phys. Rev. D 41 (1990) 374;
 V. Santos, R. V. Maluf and C. A. S. Almeida, Phys. Rev. D 93 (2016) no.8, 084047;
S. Fernando and C. Holbrook, Int. J. Theor. Phys. 45 (2006) 1630;
J. L. Blzquez-Salcedo, F. S. Khoo and J. Kunz, arXiv:1706.03262;
S. K. Chakrabarti, Gen. Rel. Grav. 39 (2007) 567;
R. Konoplya, Phys. Rev. D 71 (2005) 024038; B. Wang, C. Lin, E. Abdalla, \PLB{\bf 481}{79}{2000}; B. Wang, C. Lin, C. Molina, \PRD{\bf 70}{064025}{2004}; B. Wang, E. Abdalla, R. B. Mann, \PRD{\bf 65}{084006}{2002}; F. W. Shu and Y. Shen, \PLB{\bf 619}{340}{2005}; \PRD {\bf 70}{084046}{2004};
\JHEP{\bf 0608}{087}{2006}; J. Jing and Q. Pan, \NPB{\bf 728}{109}{2005}; \PLB{\bf 660}{13}{2008};  J. Jing, \PRD{\bf 69}{084009}{2004}; S. Chen and J. Jing,
\PLB {\bf 687}{124}{2010}; \CQG{\bf 22}{4651}{2005}; \CQG {\bf 22}{533}{2005}; \CQG {\bf 22}{2159}{2005}; S. W. Wei, Y. X. Liu, et al., \PRD{\bf 81}{104042}{2010}; B. Chen and Z. Xu,  \JHEP{\bf 0911}{091}{2009}; X. He, B. Wang and S. Chen, \PRD{\bf 79}{084005}{2009}; R. Li and J. Ren, \PRD{\bf 83}{064024}{2011}.




\bibitem{car:04}Cardoso, V., Lemos, J.P.S., Yoshida, S.: Phys. Rev. D69, 044004 (2004).


\bibitem{maeda:2006}G. Kunstatter, H. Maeda and T. Taves, \CQG{33}{105005}{2016}.


\bibitem{coll:2019}A. Colleaux, Regular black hole and cosmological spacetimes in Non-Polynomial Gravity theories. PhD thesis, University of Trento, 2019.









\end{thebibliography}
\end{document}